\documentclass[11pt,a4paper]{article}
\pdfoutput=1
\usepackage{jcappub}
\usepackage[T1]{fontenc}
\usepackage[utf8]{inputenc}
\usepackage{amssymb}

\begin{document}
\hfill {\tt DESY 16-085}

\title{Sharing but not Caring:\\Dark Matter and the Baryon Asymmetry of the Universe}

\author[1,2]{Nicolás Bernal,}
\emailAdd{nicolas@ift.unesp.br}

\author[3]{Chee Sheng Fong,}
\emailAdd{fong@if.usp.br}

\author[3,4]{Nayara Fonseca.}
\emailAdd{nayara@if.usp.br}

\affiliation[1]{ICTP South American Institute for Fundamental Research\\
                Instituto de Física Teórica, Universidade Estadual Paulista, São Paulo, Brazil}

\affiliation[2]{Institute of High Energy Physics, Austrian Academy of Sciences\\
                Nikolsdorfer Gasse 18, 1050 Vienna, Austria}

\affiliation[3]{Instituto de Física, Universidade de São Paulo, São Paulo, SP 05508-900, Brazil}

\affiliation[4]{DESY, Notkestra\ss e 85, 22607 Hamburg, Germany}

\abstract{
We consider scenarios where Dark Matter (DM) particles
carry baryon and/or lepton numbers, which can be defined if there exist operators 
connecting the dark  to the visible sector. As a result, the DM fields 
become intimately linked to the Standard Model (SM) ones and can be 
maximally asymmetric just like the ordinary  matter. In particular, 
we discuss minimal scenarios where the DM is a complex scalar or 
a Dirac fermion coupled to operators with nonzero baryon and/or lepton 
numbers, and that consist of only SM fields. We consider an initial asymmetry 
stored in either the SM or the DM sector; the main role of these operators 
is to properly {\it share} the asymmetry between the two sectors, in 
accordance with observations. After the chemical decoupling, the DM and SM sectors 
do {\it not care} about each other as there is only an ineffective communication 
between them. Once the DM mass is specified, 
the Wilson coefficients of these operators are fixed by the 
requirement of the correct transfer of the asymmetry.
We study the phenomenology of this framework at colliders, direct detection and indirect detection experiments.
In particular, the LHC phenomenology is very rich and can be tested in different channels such as the two same-sign leptons with two jets, monojet and monojet with a monolepton.
}

\maketitle

\section{Introduction}

While there is no evidence of the presence of primordial antimatter in 
the universe, the amount of primordial matter, i.e. baryons, has been determined
quite precisely from two independent observables. From the measurements of
the primordial deuterium abundance of the Big Bang Nucleosynthesis (BBN) and 
the temperature anisotropies in the Cosmic Microwave Background (CMB), the amount of baryons
as a fraction of cosmic critical energy density is determined to be 
$\Omega_{B}^{\rm{BBN}} = 0.048 \pm 0.002$~\cite{Cooke:2013cba} and 
$\Omega_{B}^{\rm{CMB}} = 0.048 \pm 0.001$~\cite{Ade:2015xua}, respectively.
In addition, the CMB measurement also yields the amount of nonbaryonic matter,
i.e. the so-called Dark Matter (DM), to be $\Omega_X = 0.258 \pm 0.008$~\cite{Ade:2015xua}. 
Given that the evidences of DM arise only from gravitational effects, it could be some 
form of exotic matter or a particle  very similar to its baryonic counterpart, 
in particular it could be {\it asymmetric}. 
The simplest asymmetric DM is either a complex scalar $\phi$ or a Dirac fermion $\psi$
uncharged under the SM gauge group. 
A single Weyl fermion is not a suitable 
candidate since it would be either massless or have a Majorana mass, meaning that it cannot 
carry an asymmetry.
The idea of an asymmetric DM giving rise to comparable DM
and baryon densities is a few decades old~\cite{Nussinov:1985xr,Roulet:1988wx,Barr:1990ca}\footnote{The connection between DM and the baryon asymmetry has also been explored in the context of {\it symmetric} DM (See e.g. Refs.~\cite{McDonald:2011zza,McDonald:2011sv,D'Eramo:2011ec,Cui:2011ab,Canetti:2012vf,Davidson:2012fn,
Bernal:2012gv,Bernal:2013bga,Racker:2014uga,Cui:2015eba}).}. 
In recent years, this idea has received a renew impetus and a plethora of the new ideas 
culminate in some recent review articles~\cite{Davoudiasl:2012uw,Petraki:2013wwa,Zurek:2013wia, Boucenna:2013wba} 
(see also~\cite{Cui:2011qe,Servant:2013uwa, Davidson:2013psa}).
Just like in a baryogenesis scenario, when Sakharov's conditions~\cite{Sakharov:1967dj}
are fulfilled, one could dynamically generate an asymmetry in $X\equiv\left\{ \phi,\,\psi\right\} $.
One can take this one step further and expect that like in the Standard
Model (SM), there are additional fast interactions among the DM sector 
which efficiently annihilate away the symmetric component ($X\bar{X}\to\cdots$), ending up with only the asymmetric part.

Up to now there is no compelling evidence 
that DM communicates with the SM via interactions other than gravity; 
and if this is all that is,  the possibilities of testing DM properties in the lab are very challenging.
 On the other hand, 
there are hints of a deeper connection between DM and the SM baryons.
For instance, their energy densities today are of similar order, which would suggest a common mechanism for the origin of the two species:
\begin{equation}
r\equiv\frac{\Omega_X}{\Omega_{B}} =\frac{Y_X^{0}\,m_X}{Y_{B_{\rm{SM}}}^{0}\,m_{n}} =\frac{\left|Y_{\Delta X}\right|\,m_X}{Y_{\Delta B_{\rm{SM}}}\,m_{n}}=5.4\,,\label{eq:ratio_DM_B}
\end{equation}
where $m_X$ and $m_{n}$ are the DM and the nucleon mass, respectively. 
Here we denote $Y_{\Delta i}\equiv Y_{i}-Y_{\bar{i}}$, where $Y_{i}=n_{i}/s$ 
is the number density of $i$ normalized by the
entropic density $s=\frac{2\pi^{2}}{45} g_{\star}T^{3}$, $g_{\star}$
is the relativistic degrees of freedom that contribute to the entropy and $T$ the temperature of the thermal bath. 
In Eq.~\eqref{eq:ratio_DM_B}, the superscript `$0$' denotes the value today, $Y_{B_\text{SM}}^{0}= (8.66 \pm 0.09) \times 10^{-11}$~\cite{Ade:2015xua} and the third 
equality is due to the assumption that both DM and baryon are {\it maximally} asymmetric.\footnote{Ref.~\cite{Graesser:2011wi} considered both the symmetric and the asymmetric DM components.}
We also denote $Y_X^{0}=\left|Y_{\Delta X}\right|$ because DM today can consist of particles $X$ or antiparticles $\bar{X}$.
Notice that the asymmetric DM scenario itself does not
justify why $r\sim 1$; further theoretical inputs 
 which relate $m_X$ to $m_n$ or $Y_{\Delta X}$ to $Y_{\Delta B_\text{SM}}$ are needed (see e.g. Ref.~\cite{Bai:2013xga,Garcia:2015toa,Farina:2015uea,Farina:2016ndq}).

In this work, we are not trying to dynamically generate $r\sim 1$, but we will rather take it as a starting point.
Our aim is to consider an Effective Field Theory (EFT) description 
where the DM and the SM sectors {\it share} a common asymmetry via interactions that were typically in equilibrium.
After the chemical decoupling between the two sectors, they barely interact i.e. {\it not caring} about each other.
In particular, 
we consider an asymmetric DM scenario in which the DM is {\it not} charged under 
the SM gauge symmetry,\footnote{DM candidates can also be the lightest neutral components of 
some $SU(2)_L$ multiplets (with zero or nonzero hypercharge). The quest to find such particles 
which are automatically stable was carried out in Ref.~\cite{Cirelli:2005uq}. 
Asymmetric DM from $SU(2)_L$ multiplets with nonzero hypercharge 
were considered in Refs.~\cite{Boucenna:2015haa,Dhen:2015wra}.} which makes its 
detection particularly challenging through SM interactions. We further assume that the DM particles 
do carry nonzero lepton and/or baryon number~\cite{Agashe:2004ci,Agashe:2004bm,Farrar:2005zd,Kaplan:2009ag, Dulaney:2010dj,Perez:2013tea,Feng:2013zda, Zhao:2014nsa, Perez:2014qfa,Fukuda:2014xqa,Ohmer:2015lxa,Fornal:2015boa, Cheung:2015mea}\,\footnote{For DM realizations within baryon and lepton number as  gauge symmetries see e.g. Refs.~\cite{FileviezPerez:2010gw, Duerr:2013dza, Duerr:2013lka, Duerr:2014wra}.} which is fixed by their coupling to 
the SM fields through higher dimensional operators of the form
\begin{eqnarray}
\frac{1}{\Lambda^{(2-p/2)N+n-4}}X^N\bar{{\cal O}}_{{\rm SM}}^{\left(n\right)}\,,
\label{eq:operator_gen}
\end{eqnarray}
where $\Lambda\equiv\Lambda'/\lambda$ with $\Lambda'$ being the effective 
scale below which the effective operators are valid and 
$\lambda$ the coupling constant between the DM and the SM sectors.
${\cal O}_{\rm SM}^{\left(n\right)}$ is a SM gauge invariant operator of
dimension $n$ consisting of {\it only} SM fields and $p=1,\,2$ for 
$X$ being a fermion or a scalar, respectively. To ensure the stability of DM particles, one needs $N \geq 2$, which can be due to specific baryonic and/or leptonic charges carried by the DM.
In this work, we consider the minimal scenario with $N=2$,
where these higher dimensional operators play the crucial role of 
distributing the asymmetry between the SM and the DM sectors. 
As it will be shown, this scenario is predictive since there is only a limited number
of possible higher dimensional operators that can be written down; for a given DM mass, their
Wilson coefficients are fixed by the requirement that the asymmetry between the SM and the DM sectors is correctly distributed to match the observations.

The mass of the DM particles can span a wide range, from few GeV up to $\sim 100$~TeV.
If $X$ carries nonzero baryon number, we have to restrict $2\,m_X \gtrsim m_{n}$ 
to prevent fast nucleon decays.
Requiring all the DM symmetric component to be annihilated away, the upper bound ($m_X \lesssim 100$~TeV) has to be imposed in order to avoid unitarity violation~\cite{Griest:1989wd,Hui:2001wy}.
The heavier the DM particles are, 
the more nonrelativistic they have to be during  chemical decoupling. This happens because it is necessary to suppress their asymmetry 
density  through a Boltzmann suppression factor to obtain the correct relic abundance (see e.g. Ref.~\cite{Buckley:2010ui}). 

\begin{figure}[t]
\centering
\includegraphics[scale=0.6]{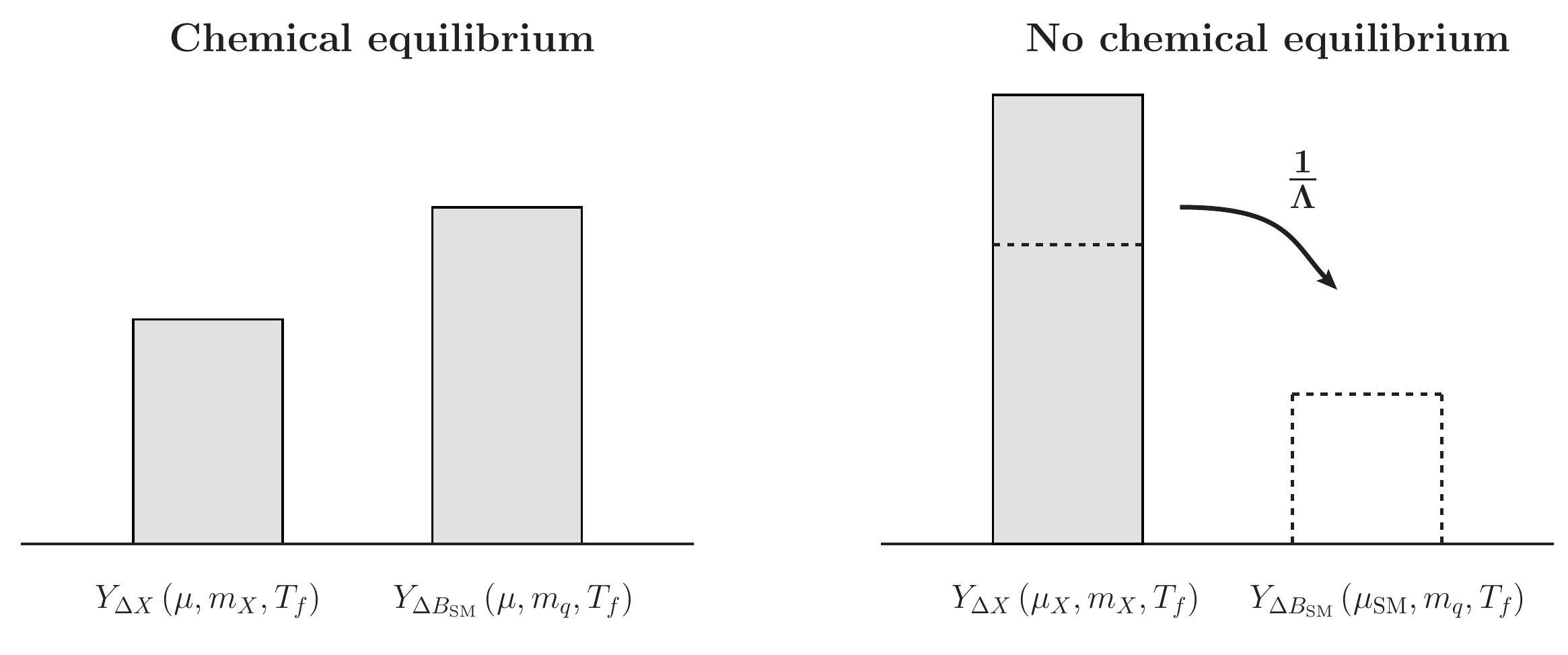}
\caption{Schematic representation of the asymmetry sharing  scenario.  Here $m_i$ refers to the mass of particle $i$.  For the chemical equilibrium case, the system is characterized by a common chemical potential $\mu$. 
For the scenario where the system never achieves chemical equilibrium, the DM and the SM sectors are described respectively 
by $\mu_X$ and $\mu_{\rm SM}$.
See text for further explanation. 
}
\label{fig:sharing}
\end{figure}

In the sharing scenario, we assume a net nonzero charge asymmetry is generated at some high scale.\footnote{This is in contrast to scenarios which consider a net zero asymmetry where the dark and visible sectors carry an equal but opposite sign asymmetry e.g. Ref.~\cite{Davoudiasl:2010am}.} Since the sharing operator~\eqref{eq:operator_gen} does not violate the charge, its role is to distribute the asymmetry among the dark and visible sectors. 
Fig.~\ref{fig:sharing} illustrates the sharing mechanism for the cases where the dark and visible sectors get into chemical equilibrium or not.  In the case where the system achieves chemical equilibrium (left panel), both asymmetries depend on the same chemical potential ($\mu$).  If the DM is nonrelativistic when the two sectors decouple at $T=T_f$,  a Boltzmann factor suppresses its number asymmetry. On the other hand, for the scenario in which the system never reaches the chemical equilibrium (right panel), the sector where the initial asymmetry resides does matter, and the asymmetries in the dark and the SM sectors are characterized by the chemical potentials $\mu_X$ and $\mu_{\rm SM}$, respectively. For instance, if the initial total asymmetry is stored in the dark sector, the amount of asymmetry transferred to the SM depends on the strength of the coupling between these two sectors, which is represented in Fig.~\ref{fig:sharing} by $1/\Lambda$.

The present  work is complementary to previous studies in the following ways: 
$i)$ our discussion is model-independent: we write down all possible effective operators and focus on the lowest dimensional ones; 
$ii)$ we cover the whole DM mass range where the effective operator description is valid; 
$iii)$ we determine the viable parameter space taking into account various
phenomenological constraints. Note that our study 
does not apply for the cases where the effective operator description is not valid
such as the scenario proposed in Ref.~\cite{Fonseca:2015rwa} where the 
mediator which links the SM and DM sectors is light, with a mass comparable to the DM one. 

The paper is organized as follows: in Section~\ref{sec2} we introduce the generalities 
of the model where DM particles carry baryon and/or lepton number. In Section~\ref{sec3}, 
we discuss in detail the transfer of the asymmetry, whether it is effective before or 
after the freeze out of the electroweak (EW) sphaleron processes. 
A number of phenomenological constraints and future detection prospects are discussed in 
Section~\ref{sec4}: DM direct and indirect detection, and the bounds coming 
from the Large Hadron Collider (LHC). Concluding remarks are presented in Section~\ref{sec5}.

\section{Baryonic and Leptonic Dark Matter}
\label{sec2}
In order to connect the DM and the SM sectors, our working assumption is that the 
DM particles carry baryon $B$ and/or lepton $L$ numbers. Of course, one could only 
define these quantum numbers if there exist operators 
which relate the DM and the SM baryons and/or leptons. 
In particular, we  consider {\it minimal} models in the sense 
that the DM particles are singlets under the SM gauge group and they
are either complex scalars or Dirac fermions.
We consider that they couple in pairs ($N=2$) to the SM, so that the operator~\eqref{eq:operator_gen} reduces to
\begin{eqnarray}
\frac{1}{\Lambda^{n-p}}XX\bar{{\cal O}}_{{\rm SM}}^{\left(n\right)}\,.\label{eq:operator}
\end{eqnarray}
To make sure that the effective operator description remains valid, 
we require $\Lambda' = \lambda\,\Lambda \gg E$, where $E$ is the characteristic energy scale being probed. 
Taking the largest coupling before perturbative expansion breaks down i.e. $\lambda = 4\pi$,
we have $\Lambda \gg E/(4\pi)$; however, one should keep in mind that the description 
could also break down earlier, for $\lambda < 4\pi$.
As we will see in more detail later, for the scenario after EW symmetry breaking,
we also impose $\Lambda > v$, where $v = \left<H\right>=174$~GeV is the Higgs 
vacuum expectation value (vev) such that this framework remains consistent.

Notice that part of our minimality criteria is to assume that 
${\cal O}_{{\rm SM}}^{\left(n\right)}$ does not contain new fields beyond the SM. 
We further assume that the operator~\eqref{eq:operator} 
preserves $B$ and $L$, which implies that
$B\left(X\right)=B\left({\cal O}_{{\rm SM}}^{\left(n\right)}\right)/2$
and $L\left(X\right)=L\left({\cal O}_{{\rm SM}}^{\left(n\right)}\right)/2$.
Assuming that the total asymmetry in a charge $q$ is preserved, 
the operator~\eqref{eq:operator} plays a role in distributing the asymmetry
among the DM and the SM sectors with 
\begin{eqnarray}
Y_{\Delta q} & \equiv & q_{X}\,Y_{\Delta X}+Y_{{\rm \Delta q}_{{\rm SM}}}
={\rm constant}\neq0\,,
\label{eq:tot_asym}
\end{eqnarray}
where $Y_{{\rm \Delta q}_{{\rm SM}}}
=\sum_{\Psi_{\rm SM}}q_{\Psi_{\rm SM}}\,Y_{\Delta {\Psi_{\rm SM}}}$ 
and $q_{i}$ is the charge of the field $i$ under $q$. 

In principle, the generation of the total asymmetry in Eq.~\eqref{eq:tot_asym} and its transfer due to the operator~\eqref{eq:operator} could happen simultaneously.
In this case, instead of being constant, the asymmetry in Eq.~\eqref{eq:tot_asym} is an evolving 
quantity which should be described by its corresponding Boltzmann Equation (BE). 
For definiteness and to be as model independent as possible,  we do not specify 
the genesis mechanism at the high scale, instead, we  assume that the asymmetry generation either 
in the DM or in the SM sector, is completed before the transfer operator~\eqref{eq:operator} 
becomes effective. As we will discuss in more detail in the next section, if the reactions induced by 
the transfer operator get into {\it chemical equilibrium} 
(i.e. proceeding faster than the expansion rate of the universe) at some point, 
the initial conditions (e.g. where the initial asymmetry resides) become irrelevant. 
On the other hand, if the transfer 
operator never gets into {\it chemical equilibrium}, 
the initial conditions do play an important role. 

The moment where the transfer of the asymmetry is efficient determines which conserved charge has to be studied.
If this transfer happens at temperatures higher than the one of the EW sphaleron processes freeze out ($T_{{\rm EWSp}}$), 
the relevant conserved charge in the SM is $q=B-L$.\footnote{Notice that $B$ and $L$ 
are violated by EW sphaleron processes but the linear combination $B-L$ remains conserved. 
Implicitly, we assume that whatever beyond the SM mechanism which violates $B-L$ 
and generates a nonzero $B-L$ asymmetry is no longer operative.}
On the other hand, if the transfer is operative at $T < T_{{\rm EWSp}}$, one can directly consider $q=B$. 
This transfer, however, should be completed before $T\sim 1$~MeV to avoid spoiling the standard BBN predictions.
Additionally, in order to make the scenario {\it predictive}, we need a further assumption:
we suppose that as in the SM, there are additional 
fast (gauge-like) interactions among the DM sector which  efficiently 
annihilate away the DM symmetric component ($X\bar{X}\to...$) and one ends up 
with only its asymmetric component.\,\footnote{The typical annihilation freeze out temperature is about $T \sim m_X/20$ while to avoid nucleon decay, the lowest DM mass we consider is about 2 GeV. Hence the lowest freeze out temperature in our scenario is about 0.1 GeV and this will not affect BBN which takes place at much lower temperature, around MeV.
If the annihilation products are some light dark particles, due to pure redshift in their temperature after decoupling, the contribution to the dark radiation during BBN or later can be estimated from the ratio of relativistic degrees of freedom $(g_{\rm BBN}/g_{\rm decoupling})^{4/3} \lesssim 0.1$ which is allowed by current Planck observations \cite{Ade:2015xua}.} Without this assumption, one could still 
study the model but it strays away from the philosophy of this work, since 
the connection between the DM and the SM through the asymmetry as in Eq.~\eqref{eq:ratio_DM_B} is lost. 
Under these considerations, the value of the conserved asymmetry $Y_{\Delta q}$ is fixed by Eqs.~\eqref{eq:ratio_DM_B} and~\eqref{eq:tot_asym}.
For instance, for $q=B-L$ and $q=B$ one has, respectively
\begin{eqnarray}
Y_{\Delta\left(B-L\right)} 
& = & Y_{\Delta\left(B-L\right)}^{0} = \left|\left(B-L\right)_{X}\right|
Y_{X}^{0}+Y_{\left(B-L\right)_{{\rm SM}}}^{0}\nonumber \\
 & = & \left[\left|\left(B-L\right)_{X}\right|r\,\frac{m_n}{m_{X}}+ \kappa \right]Y_{B_{{\rm SM}}}^{0}
\label{eq:YB-Ltot}
\end{eqnarray}
and
\begin{eqnarray}
Y_{\Delta B} & = & Y_{\Delta B}^{0}=\left|B_{X}\right|Y_{X}^{0}+Y_{B_{{\rm SM}}}^{0}\nonumber \\
 & = & \left[\left|B_{X}\right|r\,\frac{m_n}{m_{X}}+1\right]Y_{B_{{\rm SM}}}^{0}\,,
\label{eq:YBtot}
\end{eqnarray}
where $\kappa$ is an order one coefficient that relates $Y_{\left(B-L\right)_\text{SM}}^0$ and $Y_{B_\text{SM}}^0$ and which depends on the relativistic
degrees of freedom at $T_{{\rm EWSp}}$ (e.g. $\kappa=\frac{79}{28}$ if the EW sphaleron processes freeze out before EW phase transition and assuming only the SM degrees of freedom~\cite{Harvey:1990qw}). 
Let us stress that, since the nucleon mass $m_{n}\sim 1$~GeV, the SM baryon asymmetry $Y_{B_{{\rm SM}}}^{0}$
and the ratio of energy densities $r$ are fixed by observations; hence, for a given DM mass $m_{X}$, the total asymmetry (either $Y_{\Delta B}$
or $Y_{\Delta\left(B-L\right)}$) is also fixed.

In order to obtain the correct $Y_{B_{{\rm SM}}}^{0}$, one has to determine the value of $\Lambda$ for each DM mass.
To do so, one needs to track the transfer of the asymmetry from one sector to the other by solving numerically the BE (see Appendix~\ref{app:A}):
\begin{eqnarray}
\dot{Y}_{\Delta X}   & = & 
-2 \sum_{i,j,...}\left[\gamma\left(XX\leftrightarrow ij\cdots\right)
+ \gamma\left(X\bar{i}\leftrightarrow\bar{X}j\cdots\right)\right] \nonumber  \\
&  &\qquad \times \left[2\frac{Y_{\Delta X}}{g_X\, \zeta_X\, Y_0} 
- \left(\frac{Y_{\Delta i}}{g_{i}\,\zeta_i\, Y_0}
+ \frac{Y_{\Delta j}}{g_{j}\, \zeta_j\, Y_0}+...\right)\right], \label{eq:YDeltaX} 
\end{eqnarray}
where $\dot{Y}_{i}\equiv s\,H\,z\,\frac{dY_{i}}{dz}$, 
$z\equiv m_X/T$, $H=1.66\sqrt{g_{\star}}\frac{m_{X}^{2}}{M_{{\rm Pl}}\,z^{2}}$ is the Hubble expansion rate,
$M_{{\rm Pl}}$ is the Planck mass and
$Y_{0}\equiv \frac{15}{8\pi^2\,g_{\star}}$.
$\gamma (a b \leftrightarrow i j\cdots)$ 
is the thermally averaged reaction density for the scattering $a  b \leftrightarrow i j\cdots$ . 
For the particle $i$, $g_i$ denotes the corresponding degrees of freedom  while the statistical function $\zeta_i$ is given by 
\begin{equation}
\zeta_{i} \equiv \frac{6}{\pi^{2}}\int_{z_{i}}^{\infty}dx\,x\,
\sqrt{x^{2}-z_{i}^{2}}\,\frac{e^{x}}{\left(e^{x}\pm1\right)^{2}}\,,\label{eq:zeta}
\end{equation}
where $z_i \equiv m_i/T$ and the $+\,(-)$ sign corresponds to a fermionic (bosonic) field $i$.
For  relativistic particles ($z_i\ll 1$), $\zeta_i \sim 1\, (2)$.
Notice that from Eq.~\eqref{eq:tot_asym}, we have 
$\dot{Y}_{{\rm \Delta q}_{{\rm SM}}}=-q_{X}\dot{Y}_{\Delta X}$, which reflects the conservation of the charge $q$.
Hence, the symmetry of the system allows to describe the dynamics of the asymmetries using a single BE for either 
$Y_{{\rm \Delta q}_{{\rm SM}}}$ or $Y_{\Delta X}$: all the asymmetries on the right hand side of Eq.~\eqref{eq:YDeltaX}
can be written only in terms of  $Y_{{\rm \Delta q}_{{\rm SM}}}$ or $Y_{\Delta X}$~\cite{Fong:2015vna}.
Once the BE~\eqref{eq:YDeltaX} is solved and the valued of $\Lambda$ (for each $m_X$) is determined, the operator~\eqref{eq:operator} is completely fixed and  one
 can
duly calculate its phenomenological consequences which 
are discussed in detail in Section~\ref{sec4}.
In the next section, we will consider different scenarios for the sharing of the asymmetries.

\section{Scenarios for the Sharing of the Asymmetries}
\label{sec3}

In this study, we discuss two different scenarios for the transfer processes. 
In the first scenario, we consider the situation where the operator~\eqref{eq:operator} is operative and then freezes out at $T_f$, before the
EW sphaleron processes freeze out at $T_{{\rm EWSp}}$, i.e. $T_f > T_{{\rm EWSp}}$.\footnote{There could be in-between cases where the operator Eq.~\eqref{eq:operator} 
can be operative at $T > T_{{\rm EWSp}}$ and then freezes out at $T_f < T_{{\rm EWSp}}$ , but in that scenario, one loses part of the predictive power 
of the framework because the symmetries which relate the DM and the SM sectors 
are not longer the same and the relation spelled out in Eq.~\eqref{eq:tot_asym} (or 
more specifically in Eq.~\eqref{eq:YB-Ltot}) no longer applies. 
Hence we restrict the analysis to $T_f > T_{{\rm EWSp}}$.} In this regime, the  initial temperature $T_i$, defined to be the temperature when the total asymmetry generation is completed, depends on the unspecified UV completion of the model which the EFT cannot describe.  
In fact, there are  solutions that strongly depend on the initial conditions 
(and in particular on $T_i$) and they correspond to cases where the dynamics is UV-dominated.  
Hence, we only consider solutions which do not depend on $T_i$, i.e. those that achieve chemical equilibrium.
 
In the second scenario, we consider the situation where the operator Eq.~\eqref{eq:operator} 
is only operative after the EW sphaleron processes freeze out. In particular, 
the initial temperature is taken to be $T_i = T_{{\rm EWSp}}$, which we fix to $T_{{\rm EWSp}} = 132$~GeV~\cite{D'Onofrio:2014kta} and also for simplicity take this temperature to be the EW symmetry breaking scale.
In this case,
we have a well-defined initial temperature and can also entertain solutions in 
which the reactions induced by Eq.~\eqref{eq:operator} never reach the chemical 
equilibrium.

\subsection{Before the Electroweak Sphaleron Processes Freeze Out}
\label{sec:before_EW}

Here we consider the scenario where the operator Eq.~\eqref{eq:operator} 
is relevant before the EW sphaleron processes freeze out. 
In this case, the relevant symmetry of the SM is $B-L$. 
Our minimality criteria is to consider the SM gauge invariant operators 
consist of only the SM fields but carry nonzero $B-L$. 
Then, the lowest dimensional realization of the operator ${\cal O}_{{\rm SM}}^{\left(n\right)}$
corresponds to $n=5$~\cite{Weinberg:1979sa,Weinberg:1980bf}\footnote{In the following, 
the 2-component Weyl spinor notation is used. Notice that the operator
$\epsilon_{ij} \epsilon_{kl} \left(\ell_{L_\alpha}^{i} \ell_{L_\beta}^{j} \right) H^k H^l = 
{\cal O}^{\left(5\right)}_{\alpha\beta} - {\cal O}^{\left(5\right)}_{\beta\alpha}$ 
and hence it is not independent.}:
\begin{equation}
{\cal O}^{\left(5\right)}_{\alpha\beta} 
 =  \epsilon_{ik} \epsilon_{jl} 
\left(\ell_{L_\alpha}^{i} \ell_{L_\beta}^{j} \right) H^k H^l, \label{eq:before_EW_op}
\end{equation}
where $\ell_L$ and $H$ are respectively the lepton and the Higgs doublets, 
$\alpha,\beta, ...$ label the family indices, whereas
$i,j,...$ the $SU(2)_L$ indices. $\epsilon_{ij}$ is the total 
antisymmetric tensor with $\epsilon_{12}=1$.
The operator ${\cal O}^{\left(5\right)}_{\alpha\beta}$  
has zero $B$ and $L$ equals to two, which fixes $B_X=0$ and $L_X=1$.

Next we derive the relation between the particle asymmetries and
the (effective) $U(1)$ symmetries of the system~\cite{Fong:2015vna}. 
Notice that while the operator~\eqref{eq:operator} with $\cal{O}_{\rm SM}$ given by 
Eq.~\eqref{eq:before_EW_op} conserves the 
total lepton number, it generally breaks the individual lepton flavor numbers.
In the case considered here, the symmetries are the $B-L$, the hypercharge $Y$ and the $X$ number.
While the former two symmetries remain exact, 
the $X$ number is approximate in the sense that as $\Lambda \to \infty$ 
(or when the reactions induced by the operator~\eqref{eq:operator} decouple),
the $X$ number becomes a conserved quantity. We assume that the total $B-L$ is fixed by Eq.~\eqref{eq:YB-Ltot} and that the hypercharge is zero. Furthermore, the relevant SM degrees of freedom are the ones 
of the unbroken EW phase, where all the fermions are massless. Besides $\ell_L$  
and $H$, one also needs to take into account the SM fields which carry nonzero chemical potentials:
the quark doublets $q_L$, the up- and down-type quark singlets ($u_R$ and $d_R$), 
and the charged lepton singlets $e_R$, where we have suppressed the family and the color indices.

Let us consider the DM particle $X$ which carries $X$ number equal to one, 
baryon minus lepton number $\Delta_X \equiv (B-L)_X$ and zero hypercharge,
while all the SM particles carry the standard 
charge assignments. Assuming that the total hypercharge remains zero $n_{\Delta Y} = 0$, 
the number asymmetries of particles per degrees of freedom 
($SU(2)_L$ and $SU(3)_c$ multiplicities) normalized over the statistical function (Eq.~\eqref{eq:zeta})
can be expressed in terms of the $B-L$ and the $X$ charge asymmetries 
($n_{\Delta (B-L)}$ and $n_{\Delta X}$) as follows~\cite{Fong:2015vna}:
\begin{equation}
\frac{n_{\Delta i}}{g_i\,\zeta_i} = 
c_i \left(n_{\Delta (B-L)} - \Delta_X\, n_{\Delta X}\right),
\label{eq:before_EW_relation}
\end{equation}
with $c_i$ given in Table~\ref{tab:before_EW}, where the family indices for quarks and leptons have been suppressed. 
\begin{table}
\centering
\begin{tabular}{|c||c|c|c|c|c|c|}
\hline 
$\boldsymbol{i}$ & $\boldsymbol{q_{L}}$ & $\boldsymbol{u_{R}}$ & $\boldsymbol{d_{R}}$ & $\boldsymbol{\ell_{L}}$ & $\boldsymbol{e_{R}}$ & $\boldsymbol{H}$\\
\hline\hline
$\boldsymbol{c_{i}}$ & $\frac{7}{237}$ & $-\frac{5}{237}$ & $\frac{19}{237}$ & $-\frac{7}{79}$ & $-\frac{3}{79}$ & $-\frac{4}{79}$\\
\hline
\end{tabular}
\caption{The coefficients relating the number asymmetries of the corresponding 
fields to the charge asymmetries, as in Eq.~\eqref{eq:before_EW_relation}.}\label{tab:before_EW}
\end{table}

For the operator~\eqref{eq:before_EW_op}, 
$\Delta_X = -1$ and the BE for $Y_{\Delta X}$ 
from Eq.~\eqref{eq:YDeltaX} reduces to 
\begin{equation}\label{BEbefore}
\dot{Y}_{\Delta X} 
  =  - 2\gamma_{\ell\ell H H}
\left[2\frac{ Y_{\Delta X}}{g_X \zeta_X Y_0}
+\frac{22}{79}\left(\frac{Y_{\Delta (B-L)}}{Y_0}
+\frac{Y_{\Delta X}}{Y_0}\right)\right],
\end{equation}  
where $\gamma_{\ell\ell H H}\equiv\gamma_{XX\to\ell\ell HH}+\gamma_{X\bar\ell\to \bar X\ell HH}+\gamma_{XH^\dagger\to\bar X\ell\ell H}$ denotes collectively the thermally averaged reaction 
densities (defined in Eq.~\eqref{eq:reaction_density_z}) resulted from the operator~\eqref{eq:operator} with $\cal{O}_{\rm SM}$ given by 
Eq.~\eqref{eq:before_EW_op}. Notice that $Y_{\Delta (B-L)}$ here is 
fixed by observation to be $Y_{\Delta (B-L)}^0$, as in Eq.~\eqref{eq:YB-Ltot}.
The relevant reduced cross sections are collected in Appendix~\ref{app:reduced_cross_section}.

\subsection{After the Electroweak Sphaleron Processes Freeze Out}
\label{sec:after_EW}

Let us consider the scenario where the transfer
is operative after the EW sphaleron processes freeze out. 
In this case, $B$ and $L$ are the effective symmetries of the system. 
Although $L$ is not of interest here, an existing lepton asymmetry 
can affect the results as it will be shown later.
We assume that the EW symmetry is already broken and therefore that the fermions and the weak gauge bosons 
are massive. In this case, we have another relevant scale namely the vev of the Higgs $v$. 
In our EFT approach with an effective scale $\Lambda$,
we should impose $\Lambda \gg v$  to make sure the whole framework remains consistent.
As before, our minimality criteria is to consider SM gauge invariant operators 
consisting of only SM fields, and carrying nonzero $B$. 
The lowest dimensional realizations of the operator ${\cal O}_{{\rm SM}}^{\left(n\right)}$ are those for $n=6$~\cite{Weinberg:1979sa,Wilczek:1979hc,Abbott:1980zj,Alonso:2014zka}:
\begin{eqnarray}
{\cal O}^{\left(6\right){\rm I}}_{\alpha\beta\delta\gamma} 
& = & \epsilon_{abc}\epsilon_{ij} \left(q_{L_\alpha}^{ia} \ell_{L_\beta}^{j} \right) 
\left(d_{R_\delta}^{b} u_{R_\gamma}^{c} \right), \label{eq:after_EW_op1}\\
{\cal O}^{\left(6\right){\rm II}}_{\alpha\beta\delta\gamma} 
& = & \epsilon_{abc}\epsilon_{ij} \left(q_{L_\alpha}^{ia} q_{L_\beta}^{jb}\right) 
\left(u_{R_\delta}^{c} e_{R_\gamma} \right), \label{eq:after_EW_op2}\\
{\cal O}^{\left(6\right){\rm III}}_{\alpha\beta\delta\gamma} 
& = & \epsilon_{abc}\epsilon_{il}\epsilon_{jk} \left(q_{L_\alpha}^{ai} q_{L_\beta}^{jb} \right) 
\left( q_{L_\delta}^{kc} \ell_{L_\delta}^{l} \right), \label{eq:after_EW_op3}\\
{\cal O}^{\left(6\right){\rm IV}}_{\alpha\beta\delta\gamma}  
& = & \epsilon_{abc}\left( d_{R_\alpha}^{a} u_{R_\beta}^{b} \right) 
\left( u_{R_\delta}^{c} e_{R_\gamma} \right), \label{eq:after_EW_op4}
\end{eqnarray}
where $a$, $b$ and $c$ denote the color indices and $\epsilon_{abc}$ is the total antisymmetric tensor.\,\footnote{If there are right-handed neutrinos $\nu_R$'s, one could also have operator of type $u_R d_R d_R \nu_R$. If $\nu_R$ has no Majorana mass term (or if $M_{\nu_R} \ll m_X$), the sharing scenario is completely analogous to the one described by the operators \eqref{eq:after_EW_op1}--\eqref{eq:after_EW_op4} (up to gauge multiplicities). For $M_{\nu_R} \gg m_X$, the sharing operator is not relevant due to Boltzmann suppression at $T \sim m_X$. If $M_{\nu_R} \sim m_X$, this will require a separate analysis to take into account the dependence on Majorana mass term and also the effect of washout of lepton asymmetry due to lepton-number-violating Majorana mass term. In any case, the qualitative effect could be captured by our later analysis as we consider different initial conditions for lepton asymmetry: from $Y_{\Delta L} = -51/28 Y_{\Delta B}$ to $Y_{\Delta L} = 0$ to take into account various degrees of erasure of the lepton asymmetry (see Sec.~\ref{s:numerical}).}
All the operators above have both $B$ and $L$ equal to one which fixes $B_X=L_X=1/2$. 
One might also consider the dimension-7 operator
${\cal O}_{\alpha\beta\delta\gamma}^{\left(7\right)} 
= \epsilon_{abc}\epsilon_{ij} \left(u_{R_\alpha}^{a} d_{R_\beta}^{b} \right) 
\left( \bar\ell_{L_\delta}^{i} d_{R_\gamma}^{c} \right) H^{j\dagger}$
which upon the EW symmetry breaking  gives rise to 
$\epsilon_{abc}\,v\,\left(u_{R_\alpha}^{a} d_{R_\beta}^{b} \right) \left( \bar\nu_{L_\delta} d_{R_\gamma}^{c} \right)$.
For consistency, we  impose $\Lambda \gg v$  such that 
 this operator (and other higher dimensional operators proportional to powers of $v/\Lambda$) 
is subdominant and then not considered any further.

As before, we derive the relation between the particle asymmetries and
the (effective) $U(1)$ symmetries of the system. 
In general, the operator~\eqref{eq:operator} conserves total lepton number
but violates individual lepton flavor numbers, and we  assume that this is the case.
For simplicity, we also assume that the EW symmetry is already broken and hence, 
one should consider the conservation of the electric charge and   
also take into account the effect of the masses of the SM fermions. 
Due to the mass terms, the chemical potential for the left- and right-handed fields become equal.  
Therefore, in the following, we will not differentiate the fermion chiralities: 
$u = u_L = u_R$ and similarly for all the other SM fermions.
The SM fields which carry nonzero chemical potentials are the up-type 
$u$ and down-type $d$ quarks, the neutrinos $\nu$, the charged leptons $e$
and the charged weak boson $W$. As before, we have suppressed the family 
and the color indices. Hence, $u$  refers to the $u$, $c$ and $t$ quarks; analogously, 
$d$  refers to $d$, $s$ and $b$; and $e$  refers to $e$, $\mu$
and $\tau$. 

Taking the total electric charge to be zero ($n_{\Delta Q} = 0$) and 
assuming the decoupling of the top quark ($\zeta_t = 0$), one can express 
the particle number asymmetries in terms of the charge asymmetries $n_{\Delta B}$,  
$n_{\Delta L}$ and $n_{\Delta X}$ as follows:
\begin{equation}
\frac{n_{\Delta i}}{g_i\, \zeta_i} = 
\frac{1}{c_0} \left(c_B^i\, n_{\Delta B} + c_L^i\, n_{\Delta L} + c_X^i\,  n_{\Delta X}\right)\,,
\label{eq:after_EW_relation}
\end{equation}
where $c_0 \equiv 6 \left[ 5 + (2 + \zeta_b + \zeta_c + \zeta_s) \frac{3\zeta_{W}}{2}
+ (4+3\zeta_c)\zeta_b + (4+3\zeta_s)\zeta_c + 4\zeta_s \right]$,
$c_X^i = - B_X\, c_B^i - L_X\, c_L^i$, and $c_B^i$ and $c_L^i$ are given in 
Table~\ref{tab:after_EW}. 
\begin{table}
\begin{tabular}{|c||c|c|}
\hline 
$\boldsymbol{i}$ & $\boldsymbol{c_{B}^{i}}$ & $\boldsymbol{c_{L}^{i}}$\tabularnewline
\hline 
\hline 
$\boldsymbol{u}$ & $3\left(2+\zeta_{b}+\zeta_{s}+\frac{3\zeta_{W}}{2}\right)$ & $2\left(1+\zeta_{b}+\zeta_{s}\right)$\tabularnewline
\hline 
$\boldsymbol{d}$ & $3\left(3+2\zeta_{c}+\frac{3\zeta_{W}}{2}\right)$ & $-2\left(1+\zeta_{c}\right)$\tabularnewline
\hline 
$\boldsymbol{\nu}$ & $-2\left(1-\zeta_{b}+2\zeta_{c}-\zeta_{s}\right)$ & $2\left[3+\left(2+\zeta_{b}+\zeta_{c}+\zeta_{s}\right)\frac{\zeta_{W}}{2}+\left(2+\zeta_{c}\right)\zeta_{b}+\left(2+\zeta_{s}\right)\zeta_{c}+2\zeta_{s}\right]$\tabularnewline
\hline 
$\boldsymbol{e}$ & $1-\zeta_{b}+2\zeta_{c}-\zeta_{s}$ & $2\left[\left(2+\zeta_{b}+\zeta_{c}+\zeta_{s}\right)\frac{\zeta_{W}}{2}+\left(1+\zeta_{c}\right)\left(1+\zeta_{b}+\zeta_{s}\right)\right]$\tabularnewline
\hline 
$\boldsymbol{W}$ & $3\left(1-\zeta_{b}+2\zeta_{c}-\zeta_{s}\right)$ & $-2\left(2+\zeta_{b}+\zeta_{c}+\zeta_{s}\right)$\tabularnewline
\hline 
\end{tabular}
\caption{The coefficients relating the number asymmetries of the corresponding 
fields to the charge asymmetries, as in Eq.~\eqref{eq:after_EW_relation}.}\label{tab:after_EW}
\end{table}

For the operators~\eqref{eq:after_EW_op1}--\eqref{eq:after_EW_op4} we have $B_X = L_X = 1/2$; 
the BE for $Y_{\Delta X}$ is
\begin{equation}\label{BEafter}
\dot{Y}_{\Delta X} 
  =  - 2\gamma_{qqq\ell}
\left[2\frac{Y_{\Delta X}}{g_X\, \zeta_X\, Y_0}
-\frac{1}{c_0\,Y_0}\left(
c_B\, Y_{\Delta B} + c_L\, Y_{\Delta L}
-\frac{1}{2}(c_B+c_L) Y_{\Delta X}
\right)\right]\,,
\end{equation}
where
\begin{eqnarray}
c_B & = & 22 
+5\zeta_s + \frac{27\zeta_W}{2}
+ 5\zeta_b + 8\zeta_c\,, \label{cb}\\
c_L & = & 2 \left[ 2 + (2+\zeta_c+\zeta_s)\frac{\zeta_W}{2}
+ \left(3+\zeta_c+\frac{\zeta_W}{2}\right)\zeta_b + (3+\zeta_c)\zeta_s \right]\,.\label{cl}
\end{eqnarray}
$\gamma_{qqq\ell}\equiv\gamma_{XX\to qqq\ell}+\gamma_{X\bar\ell\to\bar Xqqq}+\gamma_{X\bar q\to\bar Xqq\ell}$ denotes collectively the thermally averaged reaction densities
resulted from operator~\eqref{eq:operator} with $\cal{O}_{\rm SM}$ given by 
any of the operators~\eqref{eq:after_EW_op1}--\eqref{eq:after_EW_op4}.
Although in the numerical calculations of the next section we will consider 
the statistical functions $\zeta_i$ in Eq.~\eqref{eq:zeta} for $W$, $b$, $c$ and $s$,
it turns out that the only particle which might decouple during the evolution is $W$ 
and its effect is negligible. Hence, it is a good approximation 
to consider the case where these particles are fully relativistic, i.e.
$\zeta_W/2 = \zeta_b = \zeta_c = \zeta_s = 1$ with $\{c_0,\,c_B,\,c_L\}=\{228,\,67,\,30\}$. 
Notice that $Y_{\Delta B}$ here is fixed by observations 
to be $Y_{\Delta B}^0$ as in Eq.~\eqref{eq:YBtot} while the value for
$Y_{\Delta L}^0$ is model-dependent.
As before, the relevant reduced cross sections are collected in Appendix~\ref{app:reduced_cross_section}.

\subsection{Numerical Results }
\label{s:numerical}

\begin{figure}[t]
\centering
\includegraphics[height=7.5cm]{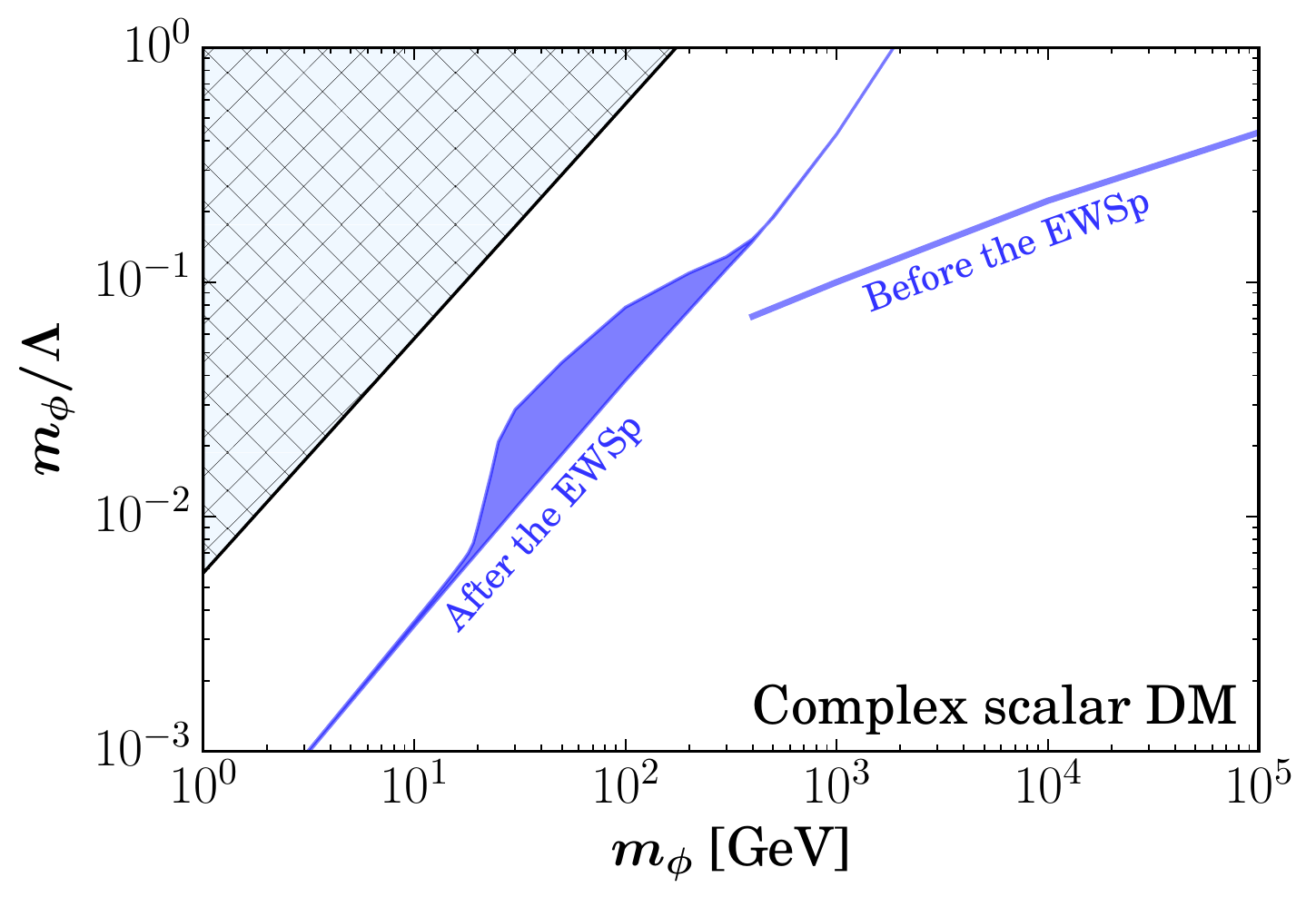}
\includegraphics[height=7.5cm]{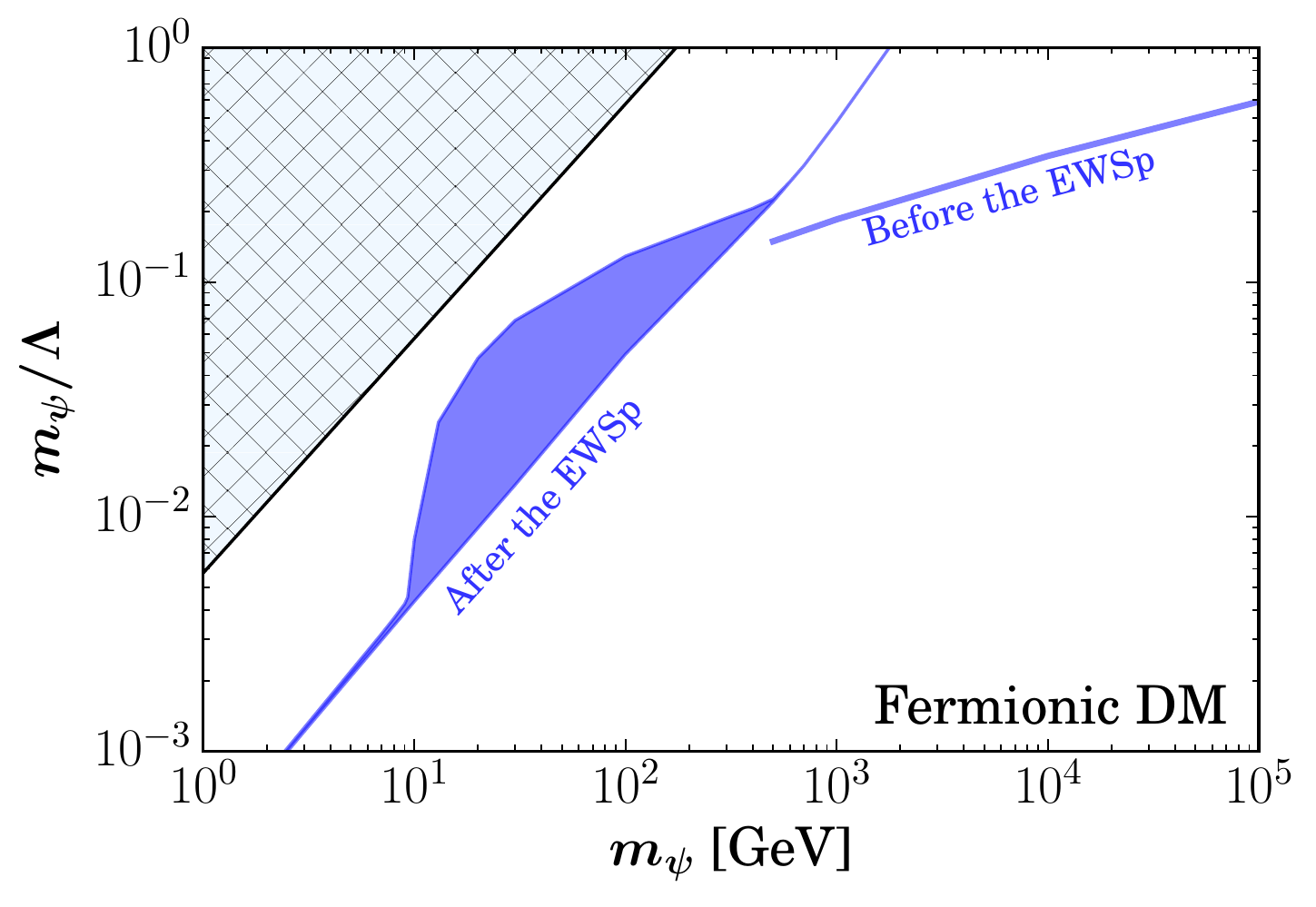}
\caption{Parameter space where the measured baryon asymmetry of the universe and the DM relic abundance can be reproduced simultaneously, for
complex scalar (upper panel) and fermionic (lower panel) DM, and for the scenarios where the transfer of the asymmetry is efficient before and after the freeze out of the EW sphalerons (see text).
In the upper left hatched regions $\Lambda$ is smaller than the Higgs vev.
}
\label{mainresults}
\end{figure}

In principle, for each DM mass $m_X$ one can solve the BE and find the appropriate value for $\Lambda$ in order to distribute the asymmetries of the two sectors such that the observed DM relic abundance and the baryon asymmetry are reproduced.
However, the results depend on the following assumptions:
\begin{enumerate}
\item[$(i)$] {\it The total asymmetry.} 
We assume that the total asymmetry is fixed by 
Eq.~\eqref{eq:YB-Ltot} or Eq.~\eqref{eq:YBtot}, for 
scenarios where the transfer of the asymmetry happens before or after the EW sphaleron processes freeze out, as described in 
the preceding section. In other words, the genesis of the asymmetry 
is already completed prior to the transfer. For the case after the 
freeze-out of the EW sphaleron processes, 
since the total lepton number is conserved, there is an extra dependence on the 
initial total lepton asymmetry. In this case, we vary 
$Y_{\Delta L}$ from $-\frac{51}{28} Y_{\Delta B}$ to 0. 
The former value is set by the EW sphaleron processes that freeze out 
before EW phase transition assuming the SM relativistic degrees of 
freedom~\cite{Harvey:1990qw}, while the latter considers the possibility of a  mechanism that erases the lepton asymmetry 
(after the freeze out of the EW sphaleron processes). This variation  is represented by the bands of solutions in Fig.~\ref{mainresults}.

\item[$(ii)$] {\it The initial distribution of the asymmetry.} 
Even if the total asymmetry is fixed by Eq.~\eqref{eq:YB-Ltot} or Eq.~\eqref{eq:YBtot}, one
has to consider whether the initial asymmetry resides 
in the DM or in the SM sector.
Let us remember that this discussion is only relevant in the case where the sharing operator does not reach the chemical equilibrium. 
If the chemical equilibrium is attained, the evolution of the system becomes independent of the initial conditions.
In this study, we assume that all the initial 
asymmetry resides either in the DM or in the SM sector. Of course, it is also possible that part 
of the initial asymmetry resides in the dark sector while the rest in the visible sector. 
However, without a specific model for the asymmetry generation, this assumption appears too 
contrived and also results in a loss of predictivity.
\end{enumerate}

In order to explore the solutions of the BE corresponding to the cases where the asymmetry is transferred before and after the freeze out of the EW sphalerons (Sections~\ref{sec:before_EW} and~\ref{sec:after_EW}, respectively),
 we consider
the two cases where $X$ is a complex scalar or a Dirac fermion.
For $X$ as a Dirac fermion, for each $\cal{O}_{\rm SM}$, there are two possible kinds of couplings,
one which involves $X_L X_L$ and the other which involves $X_R X_R$. For simplicity,
we assume both to couple equally. For definiteness, 
we further assume that DM only couples to the {\it first family} SM fermions
in Eqs.~\eqref{eq:before_EW_op} and~\eqref{eq:after_EW_op1}, and hence all the SM fermions
involved can be taken to be massless. The corresponding results for the operators~\eqref{eq:after_EW_op2}--\eqref{eq:after_EW_op4}
can be obtained by the rescaling due to different gauge multiplicities listed in
Table~\ref{tab:multiplicity_factors} (Appendix~\ref{app:reduced_cross_section}).
With the assumption of couplings only with the first family of SM fermions,
the limits from collider searches presented in the next section
are the most stringent. Furthermore, there are no flavor violating
processes. If the assumption of couplings only to the first family is relaxed,
these processes have to be taken into account. However, this introduces model dependency since 
for a complete analysis, one would need to consider a UV completion for the operators
(see for instance Ref.~\cite{Kim:2013ivd}). For the purpose of the present work,
this possibility is not considered.

In Fig.~\ref{mainresults}, we show the regions, in the plane $[m_X/\Lambda\,,m_X]$, where the measured baryon asymmetry of the universe and the DM relic abundance can be reproduced simultaneously, for complex scalar (upper panel) and fermionic (lower panel) DM. 
In the upper left hatched parts of the figure, the effective scale $\Lambda$ is smaller than the Higgs vev. This region is disregarded, 
as discussed in Section~\ref{sec:after_EW}.
Moreover, during the freeze out the relevant energy scale is $E\sim 2\,m_X$ 
and the perturbative constraint yields $m_X/\Lambda\ll 2\pi$, 
taking the coupling to be $4\pi$. In order to be conservative, 
in Fig.~\ref{mainresults} we cut off at $m_X/\Lambda = 1$.
For the numerical analysis we are fixing $r=\Omega_X/\Omega_B=5.4$ and $Y_{\Delta B_\text{SM}}=9\cdot 10^{-11}$. 
The two scenarios discussed previously in Section~\ref{sec3}, namely when the transfer of the 
asymmetry is efficient before and after the freeze out of the EW sphalerons, are depicted in the figure.

For the scenario where the asymmetry sharing takes place before the EW sphaleron processes freeze out, we only consider the solutions when the system achieves chemical equilibrium during its evolution. As $m_X$ increases, to obtain the right relic abundance, the number asymmetry of $X$ needs to decrease. This can be achieved by increasing the ratio $m_X/\Lambda$ such that the chemical decoupling happens at a later time when the number density is more Boltzmann suppressed. Notice that the increase in $m_X/\Lambda$ is quite mild due to strong Boltzmann suppression from the increase in $m_X$.
Note that for this case to work the DM has to be heavier than $\sim 500$~GeV, otherwise the freeze out occurs after the EW sphaleron freeze out.

For the scenario where the transfer happens after the EW sphaleron processes freeze out,  
the upper and lower bounds of the bands represent the two different initial total lepton asymmetries 
discussed previously: $Y_{\Delta L}=0$ and $Y_{\Delta L}=-\frac{51}{28} Y_{\Delta B}$, respectively.
For $Y_{\Delta L}=-\frac{51}{28} Y_{\Delta B}$, the system never reaches
chemical equilibrium,
which implies a dependence on the initial conditions: the initial asymmetry can reside either in the DM or in the SM sector.
Let us first consider the scenario where all the asymmetry is stored in the dark sector.
The fact that the system never reaches chemical equilibrium implies that there is no Boltzmann suppression.  
Hence, as $m_X$ increases, the ratio $m_X/\Lambda$ has to be strongly enhanced in order to deplete the number asymmetry of the DM (or equivalently, increase the transfer of the number asymmetry to the SM sector) to obtain the right relic abundance.
In this case, the increase in $m_X/\Lambda$ has to be quite steep as $m_X$ increases. 
Next we consider the case where the initial asymmetry is stored in the SM sector. 
We found that this scenario is not viable for the following reason. 
In fact, the transfer from the visible to the dark sector increases the DM asymmetry, 
but its value cannot reach the observed one which is higher than the chemical equilibrium value.

As we raise $Y_{\Delta L} \to 0$, in the mass range 
$10$~GeV $\lesssim m_X \lesssim 500$~GeV the system gets into 
chemical equilibrium. In this regime, 
the results depend quite significantly on the assumed initial lepton asymmetry. For $m_X \lesssim 10$~GeV 
or $m_X \gtrsim 500$~GeV, the system is not able to reach chemical equilibrium and 
interestingly, the results become rather independent of the existing lepton asymmetry. 
This can be understood from the following: in this regime, 
the first term of the right hand side of the BE~\eqref{BEafter}, 
which is independent on $Y_{\Delta L}$, becomes the dominant one; 
while the other terms, including the one that depends on $Y_{\Delta L}$, are subdominant.
In this case where the transfer happens after the EW sphaleron processes freeze out, the DM spans a large range of masses from few GeV to $\sim 2$~TeV.

\section{Phenomenological Constraints}\label{sec4}

Now we discuss several constraints and experimental 
opportunities due to the following two realizations of the operator~\eqref{eq:operator}:
\begin{eqnarray}
\frac{1}{\Lambda^{5-p}}XX \bar{\cal O}^{(5)} 
&=& \frac{1}{\Lambda^{5-p}}XX \epsilon_{ik} \epsilon_{jl} 
\left(\bar\ell_{L}^{i} \bar\ell_{L}^{j} \right) H^{k\dagger} H^{l\dagger}\,, \label{eq:op5}\\
\frac{1}{\Lambda^{6-p}}XX \bar{\cal O}^{\rm (6)I} 
&=& \frac{1}{\Lambda^{6-p}}XX \epsilon_{abc}\epsilon_{ij} \left(\bar q_{L}^{ia} \bar\ell_{L}^{j} \right) 
\left(\bar d_{R}^{b} \bar u_{R}^{c} \right)\, , \label{eq:op6}
\end{eqnarray}
where we have considered the coupling only to the first family SM fermion (the indices are dropped) 
and selected only ${\cal O}^{\rm (6)I}$ among the four operators from
Eqs.~\eqref{eq:after_EW_op1}--\eqref{eq:after_EW_op4}, as it has been done
in the previous section. As we mentioned before, the results for the operators~\eqref{eq:after_EW_op2}--\eqref{eq:after_EW_op4} 
can  be obtained through the rescaling due to the different gauge multiplicities as  specified in 
Table~\ref{tab:multiplicity_factors} (Appendix~\ref{app:reduced_cross_section}); though the 
phenomenological signatures can be different (for instance, in some cases they involve only charged leptons while in others only neutrinos).

Furthermore, let  us note that the requirement of fast $X\bar{X}$ annihilations in general gives rise to both DM direct detection and collider signatures which, however, are model-dependent.
On the other hand, this kind of processes do not contribute to the DM indirect detection because there is only either $X$ or $\bar{X}$ today (i.e. DM is maximally asymmetric).
Since the present shared asymmetry scenario cannot restrict this type of operators, we will not consider them further besides assuming that they are efficient enough to remove the DM symmetric component.

\subsection{Collider}\label{sec:collider}

\begin{figure}[t]
\centering
\includegraphics[height=5.3cm]{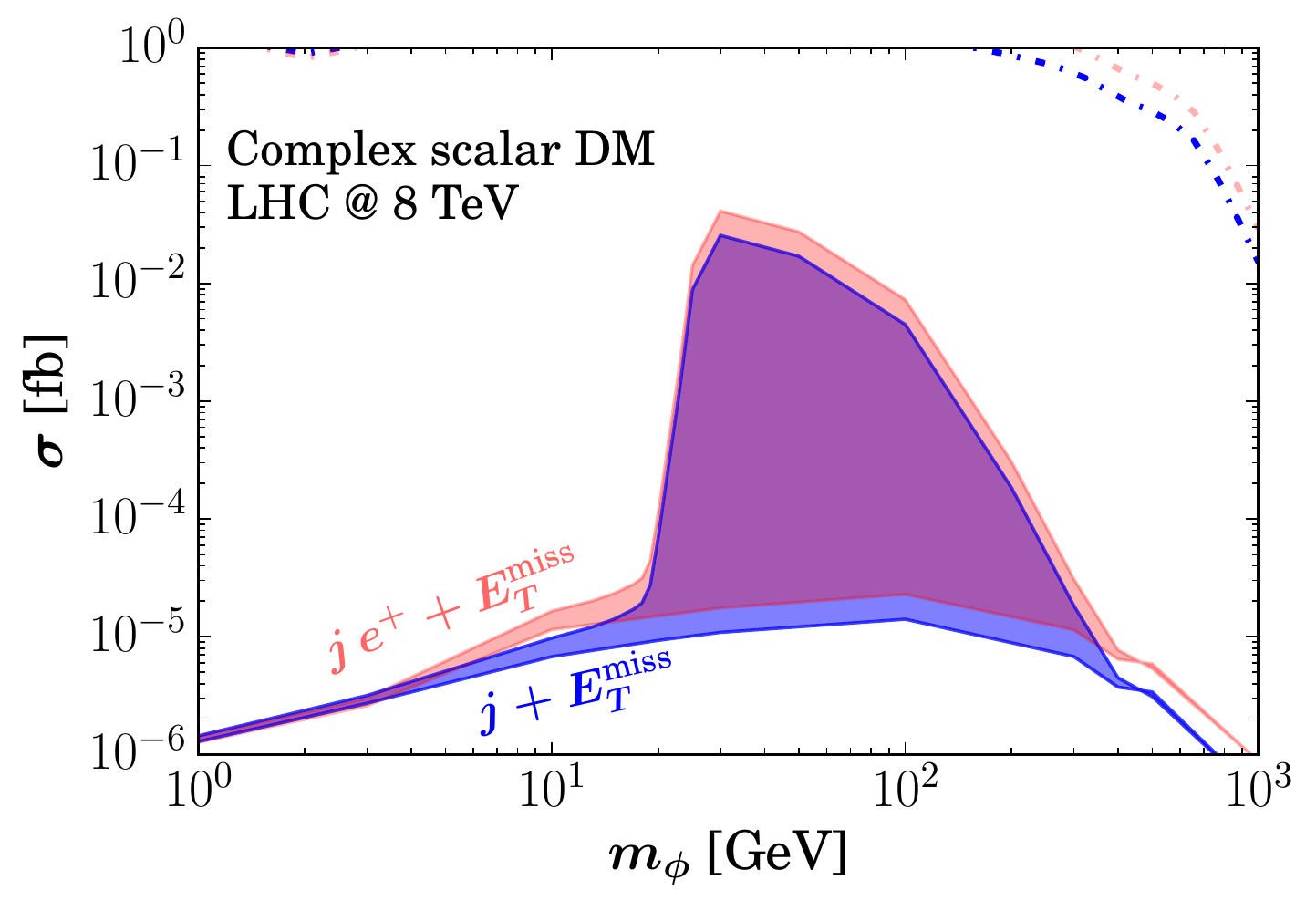}
\includegraphics[height=5.3cm]{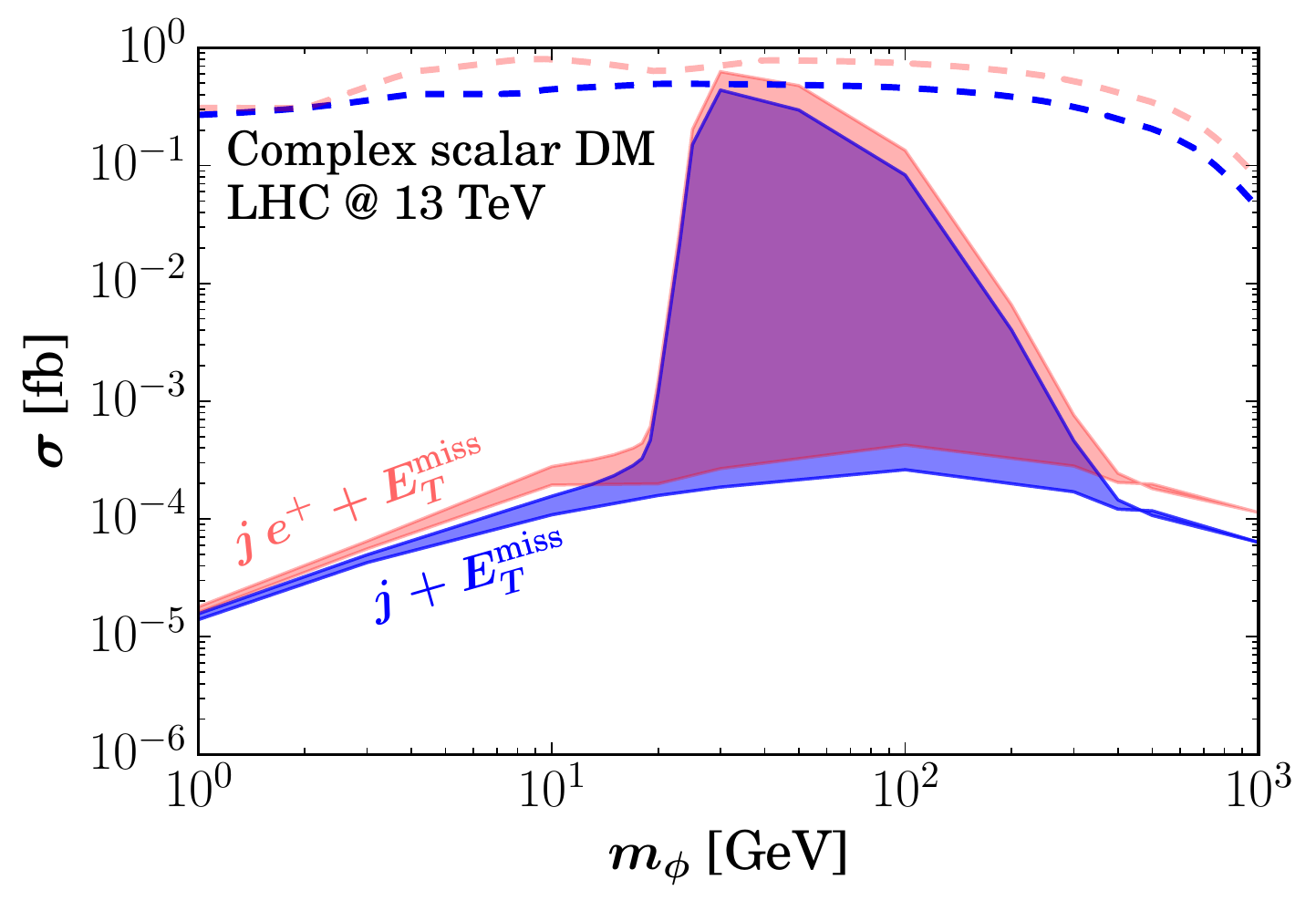}
\includegraphics[height=5.3cm]{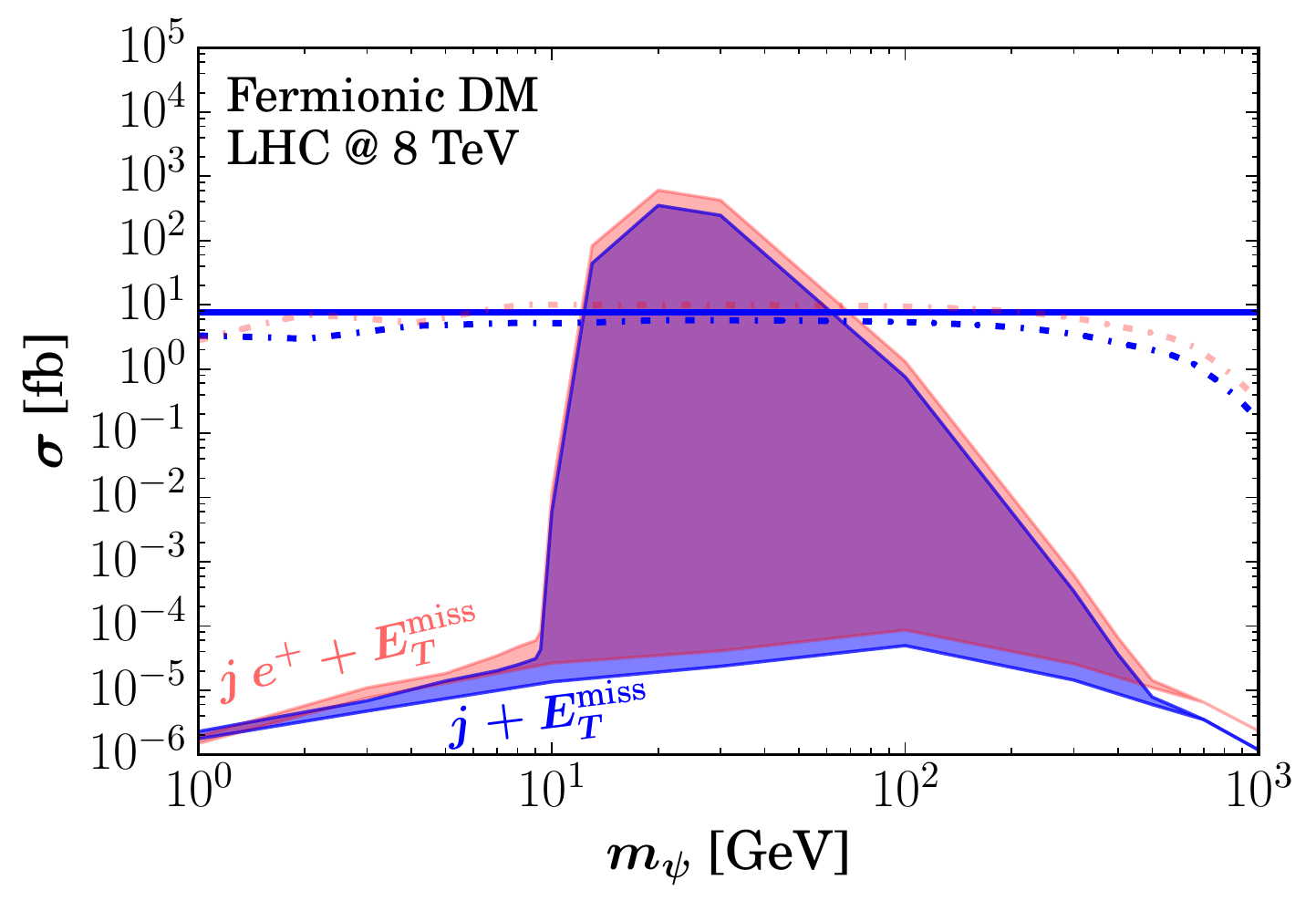}
\includegraphics[height=5.3cm]{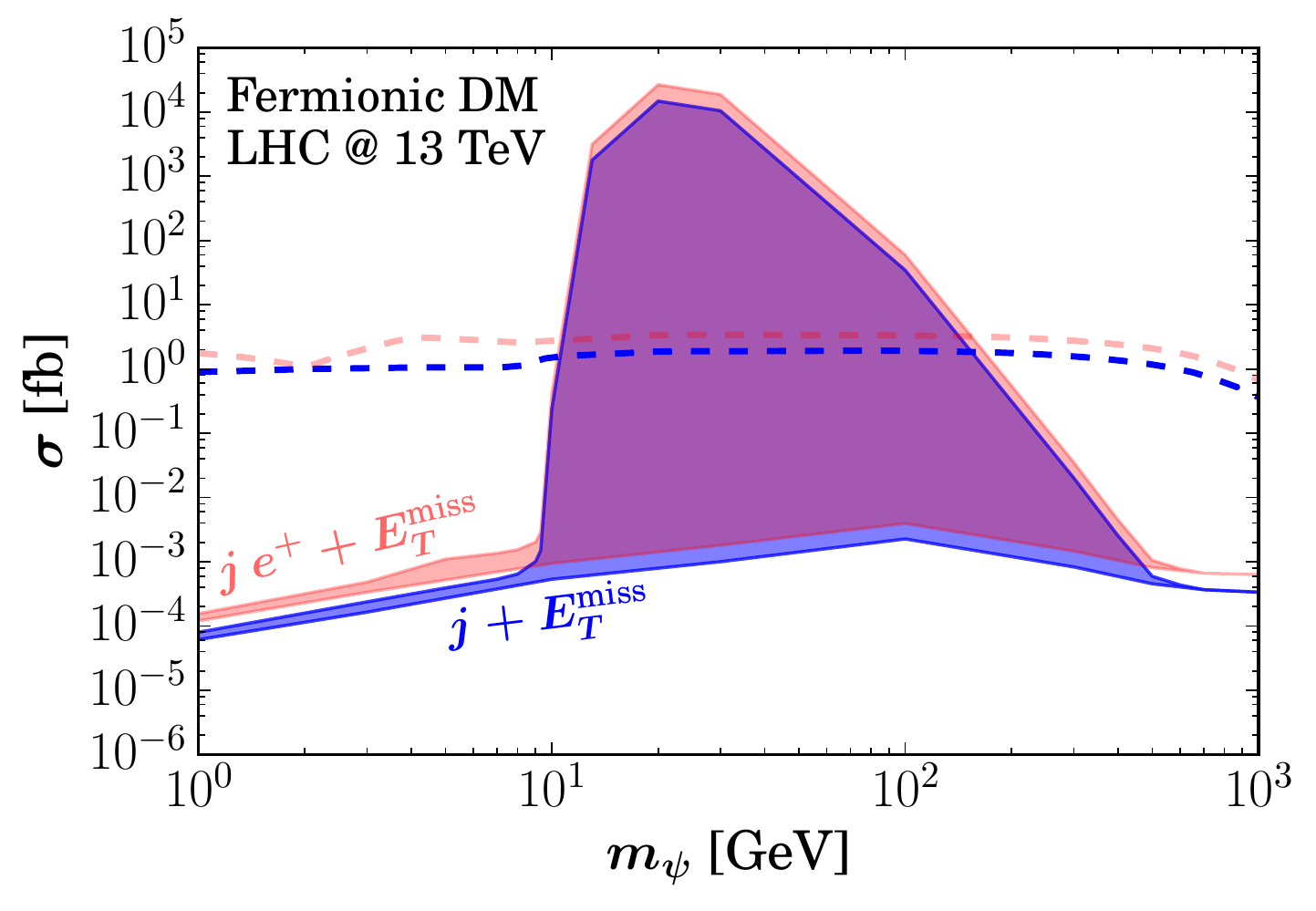}
\caption{Inclusive cross sections for monojet and missing transverse energy with (light pink) and without (blue) monolepton for scalar (first row) and fermionic (second row) DM at the LHC, for a center of mass energy of 8~TeV (first column) and 13~TeV (second column).
The blue solid line corresponds to the ATLAS exclusion limit on monojet searches; the dashed lines to the conservative limits for the break down of the EFT description (see text).
}
\label{LHC}
\end{figure}

The operator~\eqref{eq:op5} can lead to measurable signatures at colliders.
At the LHC, one can have the vector boson scattering with production of two same-sign leptons together with two jets and missing transverse energy: $pp \to W^\pm W^\pm jj$ and from our operator we will have $W^+ W^+\to e^+ e^+ XX$ and the conjugate process.
There are dedicated searches at ATLAS~\cite{Aad:2014zda,Aad:2016tuk} and CMS~\cite{Khachatryan:2014sta,Khachatryan:2016kod} that can be used in order to constraint this scenario.  The study of this process will be included  in an upcoming work~\cite{projectLHC}.
Furthermore, one could also have  $e^- e^- \to W^- W^- XX$ and the conjugate process, however, this  requires an electron-electron or a positron-positron
lepton collider, which will not be available in the near future.
 
On the other hand, the operator~\eqref{eq:op6} gives rise to two types of signatures 
at the LHC:
\begin{itemize}
\item[$(a)$] monojet with missing energy;
\item[$(b)$] monojet plus monolepton and missing energy.
\end{itemize}
There are several features 
due to that operator that we would like to highlight. 
First, for processes involving a charged lepton in the final state, we can 
further distinguish between $p p \to j\,  e^+ + E_T^{\rm miss}$ and $p p \to j\, e^- + E_T^{\rm miss}$. 
In particular, the production cross section of the latter at the LHC is 
about two to three orders magnitude more suppressed than the former, purely due to 
scarcity of antiquarks in the proton. This type of asymmetry is a distinguished feature 
of our scenario. Hence, we will only focus on the dominant process $p p \to j\, e^+ + E_T^{\rm miss}$.
Secondly, due to the steep energy dependence of the cross section of our operator 
$\sigma \propto E^{2(5-p)}/\Lambda^{2(6-p)}$, the LHC 
will be more sensitive than direct and indirect searches. 
For fermionic DM ($p=1$), the production cross section at the LHC is further enhanced. This can be understood as follows: equating the cross sections for fermion and scalar DM during the asymmetry transfer ($E \sim m_X$), we have $\Lambda_{\rm fermion}^{10} \sim  E^2 \Lambda_{\rm scalar}^{8} \sim m_X^2 \Lambda_{\rm scalar}^{8}$. Hence taking the ratio of fermion to scalar DM cross section at the LHC, we have $E^8/\Lambda_{\rm fermion}^{10} \times \Lambda_{\rm scalar}^{8}/E^6 = E^2/m_X^2$ enhancement (see Fig.~\ref{LHC}).
Thirdly, the two types of signatures $(a)$ and $(b)$ depend on the same coupling
and hence one can utilize a more sensitive channel to constraint our scenario.

In Fig.~\ref{LHC}, we show the total production cross sections for $(a)$ (blue bands) and $(b)$ (light pink bands)
at the LHC with a center of mass energy $\sqrt{s}=$ 8~TeV (left panels) and 13~TeV (right panels), using the solutions presented 
in Fig.~\ref{mainresults}. We also plot the line of $\Lambda = \sqrt{s}/(4\pi)$ to show 
the estimations below which the effective operator description could break down 
for the LHC at $\sqrt{s}=$ 8~TeV (dashed-dotted lines) and 13~TeV (dashed lines). The effective description breaks down at lower
energy if the coupling is smaller. We have assumed the most conservative value for the typically transferred energy to be the maximum value i.e. $\sqrt{s}$. Here we do not consider the unitarity bound which we expect to be of a similar order~\cite{Marciano:1989ns}.
 
\begin{figure}[t]
\centering
\includegraphics[height=5.3cm]{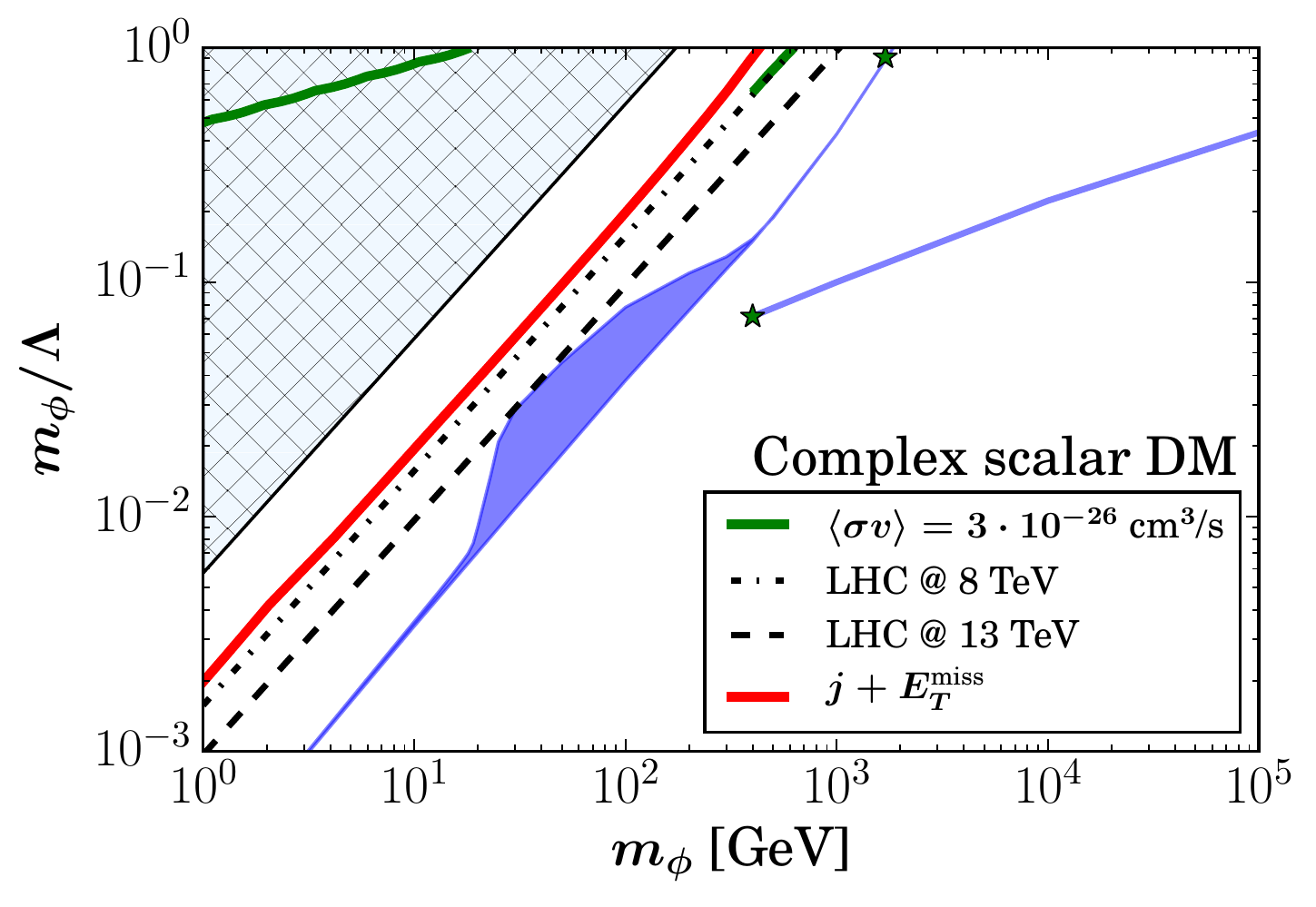}\hspace{-.2cm}
\includegraphics[height=5.3cm]{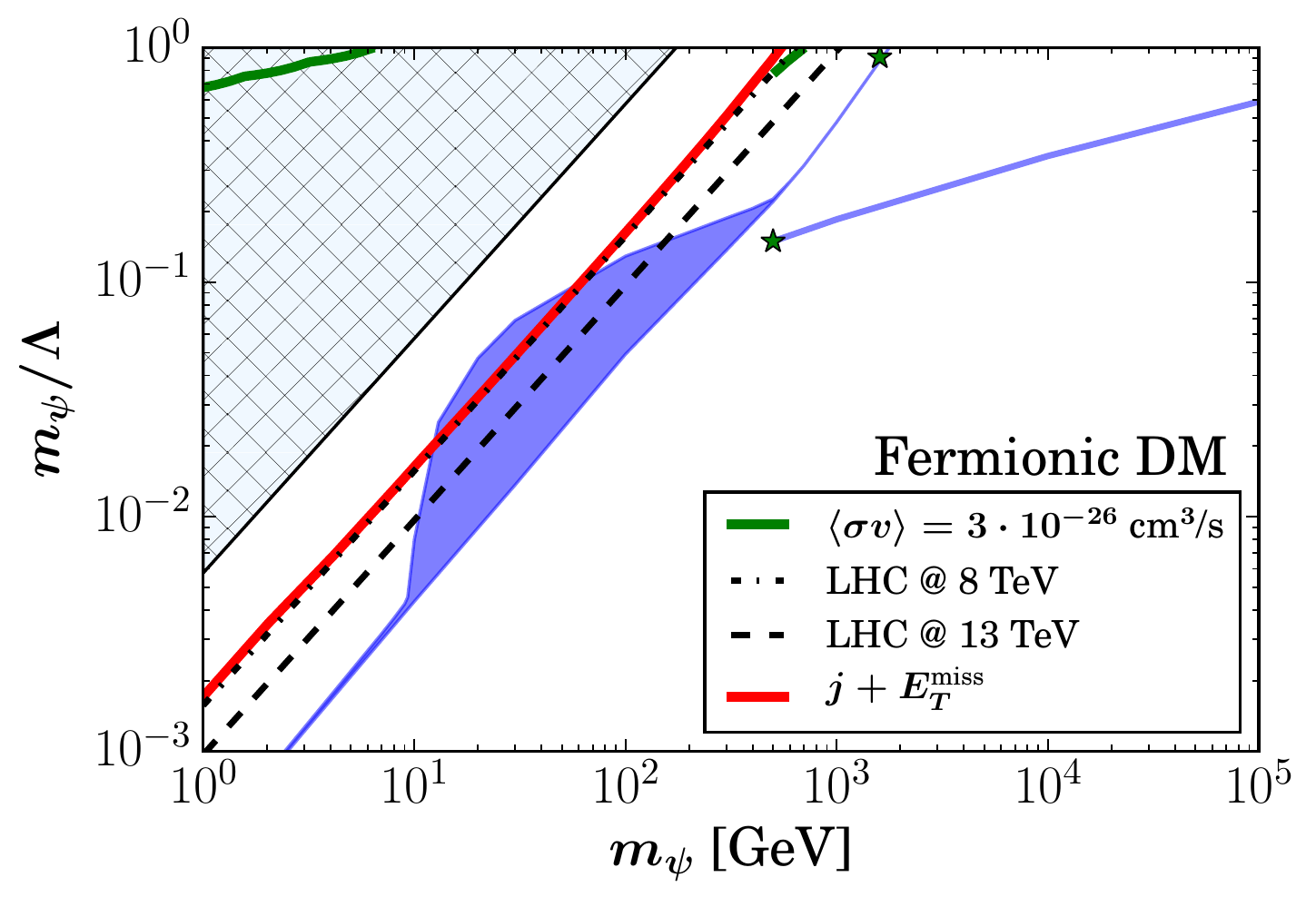}
\caption{Same as Fig.~\ref{mainresults} but adding the thermal averaged cross section  $\langle\sigma v\rangle=3\cdot 10^{-26}$~cm$^3$/s for indirect detection (solid green), the ATLAS exclusion limit from monojet searches (solid red) and the conservative limits for the break down of the EFT description for the LHC at 8 and 13~TeV (dashed-dotted and dashed lines, respectively). The stars indicate the most optimistic points for indirect detection. 
}
\label{boring}
\end{figure}

Based on monojet searches by ATLAS~\cite{Aad:2015zva} and CMS~\cite{Khachatryan:2014rra}, 
we provide an estimate on the upper bound on cross section. Using {\tt MadGraph5\_aMC@NLO}~\cite{Alwall:2014hca}, 
we estimate the efficiency times acceptance 
$\epsilon \times A$ and the most sensitive regime by imposing a cut on the final quark 
transverse momentum, which in this case, is equivalent to $E_T^{\rm miss}$. 
From the ATLAS and the CMS analysis, 
we determine the most sensitive regimes to be $E_T^{\rm miss} > 600$~GeV
and $E_T^{\rm miss} > 550$~GeV which give an upper bound on
$\sigma \times \epsilon \times A$ to be 3.8~fb 
and 7~fb at 95\% CL, respectively. Our estimation gives $\epsilon \times A \sim 0.5$ 
which implies the upper bounds $\sigma \sim $ 3.8~fb/0.5 = 7.6~fb and 7~fb/0.5 = 14~fb.
In Fig.~\ref{LHC}, the solid blue line refers to the ATLAS exclusion limit on monojet searches, which is more stringent. This bound will weaken once we do the full analysis since the efficiency will 
be lower~\cite{projectLHC}. 
In the lower left panel of Fig.~\ref{LHC} is shown that 
the monojet searches at LHC with 8~TeV (horizontal solid blue line) can already constraint a part of the parameter space of fermionic DM, corresponding to masses between $\sim 12$~GeV and $\sim 70$~GeV.
However, that region is very close to the zone where we conservatively estimate our EFT approach to break down (indicated by the dashed-dotted lines)  and a UV description would be required. In this regime, the EFT is not reliable because the LHC could in principle resolve our effective operators to reveal new heavy degrees of freedom.  However, one could still use EFT description by applying the truncation method i.e. by removing high momentum transfer events which violate the EFT to derive a weaker bound \cite{Berlin:2014cfa, Busoni:2013lha, Busoni:2014sya, Busoni:2014haa,Racco:2015dxa}. Alternatively, one could use UV complete models for analysis to obtain model dependent bounds (e.g. on new heavy states).
A more comprehensive analysis including constraints from monolepton searches 
(e.g. Ref.~\cite{Khachatryan:2014tva}) with a UV complete model will be presented in Ref.~\cite{projectLHC}. Finally, in Fig.~\ref{boring}, we translate the EFT bounds and ATLAS limit of 7.6~fb on monojet searches at 8~TeV to  the  $[m_X/\Lambda,\,m_X]$ plane.

\subsection{Dark Matter Indirect Detection}

For DM indirect detection, both operators~\eqref{eq:op5} and~\eqref{eq:op6} 
can give rise to observable astrophysical signatures. From the operator~\eqref{eq:op5}, 
the possible annihilation channels are: $XX \to \nu \nu$, 
$XX \to \nu \nu h$, $XX \to \nu \nu h h$ and $XX \to e^- e^- W^+ W^+$ and the conjugate process, where $\nu$ and $h$ are respectively the SM 
neutrino and the Higgs boson.\footnote{In our sharing scenario, 
the sign of the $B-L$ asymmetry in the DM sector is the same as the one of the SM sector (which is positive). 
Hence we will be left with only $\bar X$ today and their annihilations will necessarily contain 
antineutrinos in the final states~\cite{Fukuda:2014xqa}.} The first process dominates over the others which have additional phase space suppression. Hence, let us  consider  the thermal averaged cross section 
$\left<\sigma v\right>_{XX \to \nu \nu}$. Even for the most optimistic point 
in our parameter space $m_X = 400\,(500)$~GeV (indicated by the stars in Fig.~\ref{boring}), we have 
$\left<\sigma v\right>_{XX \to \nu \nu}\sim 5\times 10^{-32}$ cm$^3$/s 
for $X$ being a complex scalar (Dirac fermion). This value is about nine
orders of magnitude smaller than the current sensitivities of the IceCube~\cite{Aartsen:2013dxa,Aartsen:2015xej} and the ANTARES~\cite{Adrian-Martinez:2015wey} 
experiments.\footnote{In fact, 
Ref.~\cite{Queiroz:2016zwd} pointed out that for $m_X$ greater than $\sim 200$~GeV, one could 
obtain up to an order of magnitude stronger bounds utilizing the gamma-ray flux generated from EW bremsstrahlung.}

Similarly for the operator~\eqref{eq:op6}, the annihilation $XX \to$ 3 quarks + 1 lepton 
gives rise to potentially observable  fluxes of
gamma rays, neutrinos, positrons, etc. For the most optimistic case of 
$m_X\sim 2$~TeV, we determine $\left<\sigma v\right>_{XX \to \text{3 quarks + 1 lepton}} 
\sim$~few~$10^{-30}$~cm$^3$/s for both complex scalar and fermionic DM. 
This is about four orders of magnitude far from the current sensitivities of  detectors like VERITAS~\cite{Zitzer:2015eqa}, H.E.S.S.~\cite{Lefranc:2015vza} or 
MAGIC and Fermi-LAT~\cite{Ahnen:2016qkx}, which are closing in around the thermal cross section 
for standard WIMP DM $\left<\sigma v\right>_{\rm WIMP} \sim$ few $10^{-26}$ cm$^3$/s~\cite{Steigman:2012nb}. 
Hence, currently or in the near future, \ the possibilities of probing this scenario via indirect detection are very challenging.
We want to stress that this conclusion does not hold when the effective operator description does not apply
(e.g. in the proposal in Ref.~\cite{Fonseca:2015rwa}), where one can indeed have promising indirect signatures.

\subsection{Dark Matter Direct Detection}

The operator~\eqref{eq:op5} generates inelastic scatterings of the type
$X + \text{nucleon} \rightarrow \bar{X} + \text{nucleon} + 2\,\nu$ 
through a Higgs exchange. Since neutrinos can carry away momenta, the kinematic 
for this process is different from the usual 2-to-2 scattering and the sensitivity 
of the experiment should decrease.
One can do an estimation on the spin independent cross section 
$\sigma_{X-n}$ by comparing the scalar $X$ scenario to the Scalar Singlet DM model (SSDM)~\cite{McDonald:1993ex,Burgess:2000yq} (also happening via a Higgs exchange), as follows
\begin{equation}
\frac{\sigma_{X-n}}{\sigma_{{\rm SSDM}}} \sim \frac{1}{\lambda_\text{HP}^2}\left(\frac{m_{X}}{\Lambda}\right)^{6}\frac{2^{3}\pi}{2^{11}\pi^{5}}\,,
\end{equation}
where $\lambda_\text{HP}\sim \mathcal{O}\left(10^{-2}\right)$ is the Higgs-portal coupling in the SSDM. Considering the most optimistic points for $m_X = 400$~GeV and
$m_X/\Lambda =$ 0.07, one obtains $\frac{\sigma_{X-n}}{\sigma_{{\rm SSDM}}} \sim 10^{-8}$ 
(for the fermionic $X$, the ratio will be similarly suppressed).
This is much smaller than any experimental sensitivities and put us well within the regime
where the coherent neutrino scattering background becomes relevant~\cite{Billard:2013qya}.

For the operator~\eqref{eq:op6}, although we have restricted the analysis to $2\,m_X > m_n$ 
such that the nucleon decay is kinematically forbidden, 
one might have to consider the possibility of an induced nucleon 
decay (IND) as originally proposed in Ref.~\cite{Davoudiasl:2010am}. 
However, due to baryon number conservation, there is no IND in the shared asymmetry scenario 
as we explain in the following. 
First let us define the SM baryon asymmetry to be positive.
Then for the system which achieves chemical equilibrium as shown on the left panel 
of Fig.~\ref{fig:sharing}, the net baryon asymmetry has to be positive 
which also implies that the DM asymmetry has to be positive. 
From the operator~\eqref{eq:op6}, $B_X=1/2$ and hence we are left with 
only $X$ today. Also the same operator causes the IND: 
$\bar X+p\to X+e^+$, $\bar X+n\to X+\bar\nu_e$, $\bar X+p\to X+\pi^0$ and $\bar X+n\to X+\pi^0$; and since there is no 
$\bar X$ present today, these processes cannot happen.\footnote{One can 
ponder about processes like $X$ + nucleon $\to \bar{X}$ + 2 nucleons 
+ lepton, but this cannot occur due to kinematics.}
On the other hand, in the case where the DM and the SM sectors never get into 
chemical equilibrium, the initial distribution of the asymmetries become 
relevant. Since we assume that the initial total asymmetry is stored 
in either one of the sectors (e.g. on the right panel of Fig.~\ref{fig:sharing}), 
its sign has to be the same as the SM baryon asymmetry, which is positive. 
Therefore, we are left with only $X$ today and as before, IND cannot occur.
In the case where non-equilibrium dynamics generates asymmetries equal in magnitude 
and opposite in sign in the visible and DM sectors,  
the DM with opposite sign baryon number today can result in IND~\cite{Davoudiasl:2010am,Davoudiasl:2011fj,Blinov:2012hq,Huang:2013xfa,Demidov:2015bea}. 
In these scenarios, unlike in the shared asymmetry case, the transfer operators need to 
always be out of equilibrium, otherwise the asymmetry would be washed out.

\section{Conclusions}
\label{sec5}
In this work, we have considered the case where DM is a singlet under the SM gauge interactions 
but carries nonzero baryon and/or lepton numbers. In this case, the DM can be asymmetric
just like the SM baryons, and the asymmetries could be {\it shared}.
We assumed then the DM to be {\it maximally} asymmetric, and either a complex scalar or a Dirac fermion.
The DM mass spans the range between few GeV and $\sim 100$~TeV.
The connection between the dark and the visible sectors was described 
by effective operators in the context of an EFT, and it was separated in two different regimes depending on whether the transfer of the asymmetries was effective before or after the EW sphaleron processes freeze out.
The main difference between these two regimes is the following: 
before the EW sphaleron processes freeze out the relevant symmetry is $B-L$, while after the EW 
sphaleron processes freeze out $B$ and $L$ become separately the appropriate symmetries.
The leading operators consisting of only the SM fields 
come in a limited number: one dim-5 operator with $B-L$ charge and four dim-6 operators 
with $B$ charge. This feature makes the present scenario predictive in the following sense.
For a given DM mass, the total conserved asymmetry is fixed by the measurements of the ratio of energy densities $\Omega_X/\Omega_B$ and the SM baryon asymmetry $Y^0_{B_\text{SM}}$: the main role of the effective operators is to distribute the asymmetry between the visible and the dark sectors.
Furthermore, the requirement of obtaining the correct sharing (the observed DM relic abundance 
and the SM baryon asymmetry) fixes the Wilson coefficients of the operators.
Once the coefficients are fixed, one can determine the phenomenology of this scenario.

Regarding possible signatures at different facilities, we found that 
while DM indirect and direct detection are very challenging to current experimental searches, 
the LHC is already probing relevant parts of the parameter space.
This fact is due to the steep energy dependence of our operators.
The LHC phenomenology for this model is very rich and goes out of the scope of the present work.
We will dedicate a  detailed analysis in an upcoming work~\cite{projectLHC}.

\section*{Acknowledgments}
The authors want to thank Julien Billard, María Eugenia Cabrera Catalán, Oscar Éboli, André Lessa, Boris Panes 
and Alberto Tonero for helpful discussions.
NB, CSF and NF are supported by the São Paulo Research Foundation (FAPESP) under grants 2011/11973-4 \& 2013/01792-8, 2012/10995-7 \& 2013/13689-7 and 2015/15361-4, respectively.
NB is partially supported by the Spanish MINECO under Grant FPA2014-54459-P and by the `Joint Excellence in Science and Humanities' (JESH) program of the Austrian Academy of Sciences.

\appendix
\section{Derivation of the Boltzmann Equations}
\label{app:A}

First we  review the approximations made in relating 
the number density asymmetry of a particle $i$ to its chemical potential. 
The number density for $i$ can be obtained by integrating its 
phase space distribution $f_i$ over the 3-momentum as follows \cite{Kolb:1990vq}
\begin{eqnarray}
n_{i} & = & g_{i}\int\frac{d^{3}p}{\left(2\pi\right)^{3}}f_{i},\label{eq:num_den}
\end{eqnarray}
where $g_{i}$ denotes the relevant degrees of freedom (e.g. spin and gauge). 
In general, $f_i$ is a generic function that needs to be solved from the 
BE. Nevertheless, if the particle $i$ is involved in fast {\it elastic} 
scatterings (through for instance gauge interactions), the {\it kinetic} equilibrium 
is established and $f_i$ takes the equilibrium phase space distribution 
(that of Bose-Einstein or Fermi-Dirac) given by 
\begin{eqnarray}
f_{i}^{\rm eq} & = & \frac{1}{e^{\left(E-\mu_{i}\right)/T}-\eta_{i}}\,,\label{eq:dist}
\end{eqnarray}
with $\eta_{i}=1(-1)$ for $i$ boson (fermion). 
If the particle $i$ is also involved in fast {\it inelastic} scatterings with 
the corresponding gauge bosons $G$ as $i+\bar{i}\to G$, where 
$\bar{i}$ denotes the antiparticle of $i$, one has $\mu_{i}+\mu_{\bar{i}}=0 
\Rightarrow \mu_{\bar{i}}=-\mu_{i}$, since the gauge bosons have zero 
chemical potential.\footnote{This is the case for all the SM particles and 
also for the DM particle $X$ which is assumed from the outset 
to participate in  fast interactions which annihilate away the 
symmetric component.} 
In this case, the number density asymmetry of $i$ can be determined 
in terms of its chemical potential as follows
\begin{eqnarray}
n_{\Delta i} & \equiv & n_{i}-n_{\bar{i}}
=g_{i}\int\frac{d^{3}p}{\left(2\pi\right)^{3}}\left(f_{i}^{\rm eq}-f_{\bar{i}}^{\rm eq}\right)\nonumber \\
 & = & \frac{g_{i}}{2\pi^{2}}\int_{m_{i}}^{\infty}dE\,E\,\sqrt{E^{2}-m_{i}^{2}}
\left[\frac{1}{e^{\left(E-\mu_{i}\right)/T} - \eta_i}-\frac{1}{e^{\left(E+\mu_{i}\right)/T} - \eta_i}\right].\label{eq:num_den_asym}
\end{eqnarray}
For the early universe, the number density asymmetry for a particle is 
much smaller than its equilibrium number density and hence $\mu_{i}/T\ll 1$ always holds. 
However in the non-relativistic regime, since the equilibrium number density 
is Boltzmann suppressed, $\mu_{i}/T\ll 1$ might no longer hold. 
One can numerically check from Eq.~\eqref{eq:num_den_asym} that even 
if the chemical freeze-out happens as late as $T \sim m_i/25$, for $m_i \gtrsim 1$~GeV 
and imposing the correct DM asymmetry, one can still guarantee that $\mu_i/T \lesssim 0.05$.
In our asymmetry sharing scenario, such a late freeze-out 
only occur for extremely heavy $m_i$ and hence $\mu_i/T \ll 0.05$.
So one can expand Eq.~\eqref{eq:num_den_asym} in term of $\mu_{i}/T \ll 1$ and 
keep only the leading term\footnote{Notice that in the expansion, there
are only terms with odd power in $\mu_{i}/T$ as expected, since a change of sign
in the chemical potential corresponds to a change of sign in the asymmetry number density, 
as shown in Eq.~\eqref{eq:num_den_asym}. Interestingly, for massless particles, 
the series truncates at $(\mu_i/T)^3$.}
\begin{eqnarray}
n_{\Delta i} & = & \frac{T^{3}}{6}\,g_{i}\,\zeta_{i}\,\frac{\mu_{i}}{T}\,,\label{eq:num_den_asym_2}
\end{eqnarray}
where we have defined
\begin{eqnarray}
\zeta_{i} & \equiv & \frac{6}{\pi^{2}}\int_{z_{i}}^{\infty}dx\,
x\sqrt{x^{2}-z_{i}^{2}}\frac{e^{x}}{\left(e^{x} - \eta_i\right)^{2}}\,,
\end{eqnarray}
with $z_{i}\equiv m_{i}/T$. In the relativistic limit $z_{i}\ll1$,
we have $\zeta_{i} \sim 1\,(2)$ for $i$ being a fermion (boson) while in the nonrelativistic
limit $z_{i}\gg1$, we obtain $\zeta_{i}=\frac{6}{\pi^{2}}\,z_{i}^{2}\,{\cal K}_{2}(z_{i})$
for both fermions and bosons, where ${\cal K}_{2}$ is the modified Bessel function of the 
second kind of order 2. For later convenience, we further define the abundance as 
\begin{equation}
Y_a = \frac{n_a}{s},\label{eq:abundance} 
\end{equation}
where $s = \frac{2\pi^{2}}{45}g_{\star}T^{3}$ is the entropic density
and $g_{\star}$ the effective entropic degrees of freedom 
($g_{\star} = 106.75$ when all the SM particles are relativistic).
Finally, we can rewrite Eq.~\eqref{eq:num_den_asym_2} as
\begin{eqnarray}
\frac{\mu_{i}}{T} & = & 
\frac{1}{2}\frac{n_{\Delta i}}{g_{i}\, \zeta_i\, n_{0}}
= \frac{1}{2}\frac{Y_{\Delta i}}{g_{i}\,\zeta_i\,Y_{0}}\,,
\label{eq:chempot_num_den_asym}
\end{eqnarray}
where we have defined $n_{0}\equiv \frac{T^{3}}{12}$ 
and $Y_{0}\equiv \frac{n_{0}}{s} = \frac{15}{8\pi^2 g_{\star}}$.\\

Now we  proceed to derive the BE used in this work. 
In the radiation-dominated early universe described 
by the Friedmann-Lema\^{i}tre-Robertson-Walker metric, 
the evolution of the number density of particle $a$ for a 
general process $a+b+...\leftrightarrow i+j+...$ is described by~\cite{Kolb:1990vq}
\begin{eqnarray}
\dot{Y}_{a} & = & -\sum_{b,...;\,i,j,...}\left[ab...\leftrightarrow ij...\right],\label{eq:BE_Y}
\end{eqnarray}
where we have defined $\dot{Y}_{a}\equiv s\,H\,z\,\frac{dY_{a}}{dz}$ with 
$z \equiv m/T$,  $m$ some arbitrary mass scale and 
the Hubble expansion rate given by
\begin{equation}
H = \sqrt{\frac{4\pi^3 g_\star}{45}} \frac{T^2}{M_{\rm Pl}},
\end{equation}
where $M_{\rm Pl} = 1.22 \times 10^{19}$~GeV. 
The  right-handed side of Eq.~\eqref{eq:BE_Y} is the collision term 
given by
\begin{eqnarray}
\left[ab...\leftrightarrow ij...\right] & \equiv & 
\Lambda_{ab...}^{ij...}\left[\left|{\cal M}\left(ab...\to ij...\right)\right|^{2}
f_{a}f_{b}...\left(1+\eta_{i}f_{i}\right)\left(1+\eta_{j}f_{j}\right)...\right.\nonumber \\
&  &\qquad \left.-\left|{\cal M}\left(ij...\to ab...\right)\right|^{2}
f_{i}f_{j}...\left(1+\eta_{a}f_{a}\right)\left(1+\eta_{b}f_{b}\right)...\right]\,,\label{eq:reaction}
\end{eqnarray}
where
\begin{eqnarray}
\Lambda_{ab...}^{ij...} & \equiv & c_{\rm PS_I} c_{\rm PS_F} 
\int d\Pi_{a}d\Pi_{b}...d\Pi_{i}d\Pi_{j}...\left(2\pi\right)^{4}
\delta^{\left(4\right)}\left(p_{a}+p_{b}+...-p_{i}-p_{j}-...\right),\\
d\Pi_{a} & \equiv & \frac{d^{3}p_{a}}{\left(2\pi\right)^{3}2E_{a}}\,.
\end{eqnarray}
In the above, $\left|{\cal M}\left(ab...\to ij...\right)\right|^{2}$ 
is the squared amplitude {\it summed} over initial and final degrees of freedom 
(the additional factor in Wick's contraction for identical particles in the operator 
is implicitly taken into account). For later convenience, we explicitly 
write down $c_{\rm PS_I}$ and $c_{\rm PS_F}$ to denote the respective compensating factors 
when there are identical particles in the initial and final states in order to 
avoid over-counting their contributions to the phase space. 
For instance, for $n$ identical particles in the initial or final states, 
we have a factor of $c_{\rm PS_I}=1/n!$ or $c_{\rm PS_F}=1/n!$, respectively.

For a process $ab...\leftrightarrow ij...$, 
using Eq. (\ref{eq:dist}) and energy conservation $E_a + E_b + ... = E_i + E_j + ...$,
the phase space distributions with vanishing chemical potentials 
$f_i^{\rm eq,0} \equiv f_i^{\rm eq}\left(\mu_i=0\right)$ 
obey the identity 
\begin{eqnarray}
F^{{\rm eq,0}} & \equiv &
f_{a}^{{\rm eq,0}}f_{b}^{{\rm eq,0}}...
\left(1+\eta_{i}f_{i}^{{\rm eq,0}}\right)\left(1+\eta_{j}f_{j}^{{\rm eq,0}}\right)...\nonumber \\
& = & f_{i}^{{\rm eq,0}}f_{j}^{{\rm eq,0}}...\left(1+\eta_{i}f_{a}^{{\rm eq,0}}\right)
\left(1+\eta_{j}f_{b}^{{\rm eq,0}}\right)...\,.
\end{eqnarray}
Eq.~\eqref{eq:reaction} can be further rewrite as
\begin{eqnarray}
\left[ab...\leftrightarrow ij...\right] & = & \gamma\left(ab...\to ij...\right)\frac{f_{a}f_{b}...\left(1+\eta_{i}f_{i}\right)\left(1+\eta_{j}f_{j}\right)...}{F^{{\rm eq,0}}}\nonumber \\
 &  & -\gamma\left(ij...\to ab...\right)\frac{f_{i}f_{j}...\left(1+\eta_{a}f_{a}\right)\left(1+\eta_{b}f_{b}\right)...}{F^{{\rm eq,0}}},\label{eq:reaction2}
\end{eqnarray}
where we have defined the thermal averaged reaction density as
\begin{eqnarray}
\gamma\left(ab...\to ij...\right) & \equiv & 
\Lambda_{ab...}^{ij...}\left|{\cal M}\left(ab...\to ij...\right)\right|^{2}F^{{\rm eq,0}}.
\label{eq:reaction_density}
\end{eqnarray}
For the moment, the quantity above is just  schematic 
 since in Eq.~\eqref{eq:reaction2}
all the $f_i$ implicitly contain the momenta to be integrated over. 
Hence, we need the following approximations:
\begin{enumerate}
\item Ignore the corrections from Fermi-blocking and Bose-enhancement factors 
by setting 
\begin{equation}
\frac{1+\eta_{i}f_{i}}{1+\eta_{i}f_{i}^{{\rm eq,0}}} \to 1. 
\label{eq:approx1}
\end{equation}

\item If particles $i,j,...$ are in kinetic equilibrium 
(i.e. if $f_{i,j,...}$ take the form of Eq.~\eqref{eq:dist}) and 
the conditions $\mu_{i,j,...}/T\ll1$ apply, 
one can expand up to linear term in their chemical potentials as follows
\begin{equation}
\frac{f_{i}f_{j}...}{f_{i}^{{\rm eq,0}}f_{j}^{{\rm eq,0}}...} 
= 1+\frac{\mu_{i}}{T}+\frac{\mu_{j}}{T}+...
= 1 + \frac{1}{2}\frac{Y_{\Delta i}}{g_{i} \zeta_i Y_0}
+\frac{1}{2}\frac{Y_{\Delta i}}{g_{j} \zeta_j Y_0} + ...\,,
\label{eq:approx2}
\end{equation}
where in the second equality Eq.~\eqref{eq:chempot_num_den_asym} has been used.

\end{enumerate}

With the above approximations, all the energy-momentum dependent terms 
which multiply Eq.~\eqref{eq:reaction_density} in  Eq.~\eqref{eq:reaction2} 
drops out, and Eq.~\eqref{eq:reaction_density} becomes a well-defined quantity.
Now we are ready to derive the quantities we are interested in,  
which are the two types of scattering processes: 
$\left[XX\leftrightarrow ij...\right]$ and 
$\left[X\bar{i}\leftrightarrow \bar{X}j...\right]$,
where $X$ is massive while the rest of the particles are massless.
Applying the approximations~\eqref{eq:approx1} and~\eqref{eq:approx2}, 
and keeping up to first order in number asymmetry densities, one has
\begin{eqnarray}
\left[XX\leftrightarrow ij...\right] & = & \gamma\left(XX\to ij...\right)
\left( \frac{Y_{\Delta X}}{g_{X}\zeta_X Y_0}
- \frac{1}{2} \frac{Y_{\Delta i}}{g_{i}\zeta_i Y_0}
- \frac{1}{2} \frac{Y_{\Delta j}}{g_{j}\zeta_j Y_0}-...\right),\\
\left[X\bar{i}\leftrightarrow\bar{X}j...\right]
& = & \gamma\left(X\bar{i}\leftrightarrow\bar{X}j...\right)
\left( \frac{Y_{\Delta X}}{g_{X}\zeta_X Y_0}
-\frac{1}{2} \frac{Y_{\Delta i}}{g_{i}\zeta_i Y_0}
-\frac{1}{2} \frac{Y_{\Delta j}}{g_{j}\zeta_j Y_0} - ... \right).
\end{eqnarray}
For the $CP$ conjugate processes $\left[\bar X \bar X\leftrightarrow \bar i \bar j...\right]$ 
and $\left[\bar X i\leftrightarrow X \bar j...\right]$, one simply has to flip 
the signs of all number density asymmetries (i.e. chemical potentials). 
Since in this context the $CP$ violation is not relevant, 
one can consider the tree level processes where $\gamma\left(XX\to ij...\right) = 
\gamma\left(\bar X\bar X\to \bar i \bar j...\right)$
and $\gamma\left(X\bar{i}\leftrightarrow\bar{X}j...\right) = 
\gamma\left(\bar X i \leftrightarrow X \bar j...\right)$.
The BE for $Y_X$ and  $Y_{\bar{X}}$ are simply
\begin{eqnarray}
\dot{Y}_{X} & = & -2 \sum_{i,j,...}\left[XX\leftrightarrow ij...\right] 
-\sum_{i,j,...}\left(\left[X\bar{i}\leftrightarrow\bar{X}j...\right] 
- \left[\bar X i \leftrightarrow X \bar j...\right] \right), \\
\dot{Y}_{\bar X} & = & -2\sum_{i,j,...}\left[\bar X \bar X\leftrightarrow \bar i \bar j...\right]
+ \sum_{i,j,...}\left(\left[X\bar{i}\leftrightarrow\bar{X}j...\right] 
- \left[\bar X i \leftrightarrow X \bar j...\right] \right).
\end{eqnarray}
Finally, the BE for $Y_{\Sigma X}\equiv Y_X + Y_{\bar{X}}$ 
and $Y_{\Delta X}\equiv Y_X - Y_{\bar{X}}$ can be written down as follows
\begin{eqnarray}
\dot{Y}_{\Sigma X}   = & -&2 \sum_{i,j,...} \left( \left[XX\leftrightarrow ij...\right]
+ \left[\bar X \bar X\leftrightarrow \bar i \bar j...\right] \right) = 0, \label{eq:YSigmaX_app} \\ \nonumber
\dot{Y}_{\Delta X}   = & -&2 \sum_{i,j,...} \left( \left[XX\leftrightarrow ij...\right]
- \left[\bar X \bar X\leftrightarrow \bar i \bar j...\right] \right) \\ &-&2\sum_{i,j,...} \left(\left[X\bar{i}\leftrightarrow\bar{X}j...\right] 
- \left[\bar X i \leftrightarrow X \bar j...\right] \right) \nonumber \\
 = & -&2 \sum_{i,j,...}\left[\gamma\left(XX\to ij...\right)
+ \gamma\left(X\bar{i}\leftrightarrow\bar{X}j...\right)\right] \nonumber  \\
 &  & \times \left[2\frac{Y_{\Delta X}}{g_X \zeta_X Y_0} 
- \left(\frac{Y_{\Delta i}}{g_{i}\zeta_i Y_0}
+ \frac{Y_{\Delta j}}{g_{j} \zeta_j Y_0}+...\right)\right]. \label{eq:YDeltaX_app} 
\end{eqnarray}
Notice that $Y_{\Sigma X}$ is constant up to ${\cal O}(Y_{\Delta X})$. 
Since we have assumed $X$ to participate in fast scattering e.g. $X \bar X \to ...$, 
$Y_{\Sigma X} = 2\,Y_{X}^{\rm eq}$ up to ${\cal O}(Y_{\Delta X})$ 
throughout the evolution. So at the leading order in $Y_{\Delta X}$,
one only needs to solve the BE for $Y_{\Delta X}$.

\section{Reduced Cross Sections}
\label{app:reduced_cross_section}

For a 2-to-$N$ body scattering, from Eq.~\eqref{eq:reaction_density} one has that the reaction density
\begin{eqnarray}
\gamma\left(ab\to ij...\right) & \equiv & 
\Lambda_{ab}^{ij...}\left|{\cal M}\left(ab\to ij...\right)\right|^{2}f_{a}^{{\rm eq}}f_{b}^{{\rm eq}}
\left(1+\eta_{i}f_{i}^{{\rm eq}}\right)\left(1+\eta_{j}f_{j}^{{\rm eq}}\right)\dots\,.
\label{eq:reduced_cross_section_2-to-N}
\end{eqnarray}
Ignoring the Fermi-blocking and the Bose-enhancement factors, and 
taking all the phase space distribution to be Maxwell-Boltzmann, 
one has\footnote{These approximations allow to express 
Eq.~\eqref{eq:reduced_cross_section_2-to-N} as the integral of a
{\it standard} scattering cross section. Notice that however,  
the squared amplitude is summed over both the initial and the final 
degrees of freedom (no averaged over the initial degrees of freedom). 
Otherwise, one just has to evaluate Eq.~\eqref{eq:reduced_cross_section_2-to-N} directly.}
\begin{eqnarray}
\gamma\left(ab\to ij...\right) & = & c_{\rm PS_I}  
\int d\Pi_{a}\,d\Pi_{b}\,2\,s\,\beta\left(1,\frac{m_{a}^{2}}{s},\frac{m_{b}^{2}}{s}\right)
\sigma_{ab\to ij...}\left(s\right)e^{-\left(E_{a}+E_{b}\right)/T},
\label{eq:reduced_cross_section_2-to-N_MB}
\end{eqnarray}
where the cross section is
\begin{eqnarray}
\sigma_{ab\to ij...}\left(s\right) & = & 
c_{\rm PS_F} \frac{1}{2s\beta\left(1,\frac{m_{a}^{2}}{s},\frac{m_{b}^{2}}{s}\right)}
\int\left|{\cal M}\left(ab\to ij...\right)\right|^{2}d\Phi_{N},
\end{eqnarray}
with $d\Phi_{N}\equiv\left(2\pi\right)^{4}
\delta^{\left(4\right)}\left(P_{a}+P_{b}-P_{i}-P_{j}-...\right)d\Pi_{i}\,d\Pi_{j}...$
being the $N$-body phase space element and 
\begin{equation}
\beta\left(1,v,w\right)\equiv \sqrt{\left(1-v-w\right)^2-4vw}.
\end{equation}
With some algebra one can rewrite Eq.~\eqref{eq:reduced_cross_section_2-to-N_MB}
in a more convenient form in term of only the center of mass energy $\sqrt{s}$ and the temperature 
$T$ of the thermal bath as follows
\begin{eqnarray}
\gamma\left(ab\to ij...\right) & = & c_{\rm PS_I}
\frac{T}{64\pi^{4}}\int_{s_{\rm min}}^\infty ds\sqrt{s}\,\hat{\sigma}_{ab\to ij...}\left(s\right)
{\cal K}_{1}\left(\frac{\sqrt{s}}{T}\right),\label{eq:reaction_density_sim}
\end{eqnarray}
where $s_{\rm min}= \max\left[\left(m_{a}+m_{b}\right)^{2}
\left(m_{i}+m_{j}+...\right)^{2}\right]$, 
${\cal K}_1$ is the modified Bessel function of the second kind of order 1 
and the dimensionless reduced cross section is
\begin{eqnarray}
\hat{\sigma}_{ab\to ij...}\left(s\right) & \equiv & 2s\left[\beta\left(1,\frac{m_{a}^{2}}{s},\frac{m_{b}^{2}}{s}\right)\right]^{2}\sigma_{ab\to ij...}\left(s\right).\label{eq:reduced_cross_section}
\end{eqnarray}
Finally Eq.~\eqref{eq:reaction_density_sim} can be rewritten in a more convenient
form in term of $z=m/T$ with $m$ a convenient mass scale as follows
\begin{eqnarray}
\gamma\left(ab\to ij...\right) & = & c_{\rm PS_I} 
\frac{m^{4}}{64\pi^{4}z}\int_{x_{{\rm min}}}^{\infty}dx\sqrt{x}\,\hat{\sigma}_{ab\to ij...}\left(x\right)
{\cal K}_{1}\left(\sqrt{x}z\right),
\label{eq:reaction_density_z}
\end{eqnarray}
with $x\equiv s/m^{2}$ with $x_{{\rm min}}=s_{\rm min}/m^2$.

\subsection[Dark Matter as a Complex Scalar $\phi$]{Dark Matter as a Complex Scalar $\boldsymbol{\phi}$}

In the following the mass of complex scalar $\phi$ is 
taken to be $m_\phi$ while $f$ and $b$ denote respectively 
a massless Weyl fermion and a complex scalar. 
First, the relevant reduced cross sections  
(Eq.~\eqref{eq:reduced_cross_section}) that arise from an operator 
of the type $\phi \phi f f b b$ are
\begin{eqnarray}
\hat{\sigma}_{\phi \phi ffbb}\left(x\right) & = & 
c_{\rm G}\, c_{\rm S}\, c_{\rm PS_F} \frac{m_{\phi}^{6}}{\Lambda^{6}}
\beta\left(1,x^{-1},x^{-1}\right)\frac{1}{2^{11}\pi^{5}}\frac{x^{3}}{72},\\
\hat{\sigma}_{\phi b\phi bff}\left(x\right) & = & 
c_{\rm G}\, c_{\rm S}\, c_{\rm PS_F} \frac{m_{\phi}^{6}}{\Lambda^{6}}
\beta\left(1,x^{-1},0\right)\frac{1}{2^{11}\pi^{5}}F_{\phi b \phi bff}\left(x\right),\\
\hat{\sigma}_{\phi f\phi fbb}\left(x\right) & = & 
c_{\rm G}\, c_{\rm S}\, c_{\rm PS_F} \frac{m_{\phi}^{6}}{\Lambda^{6}}
\beta\left(1,x^{-1},0\right)\frac{1}{2^{11}\pi^{5}}F_{\phi f\phi fbb}\left(x\right),
\end{eqnarray}
where $x \equiv s/m_\phi^2$ and
\begin{eqnarray}
F_{\phi b\phi bff}\left(x\right) & \equiv & 
\frac{1}{72x}(x^2-1)(x^2+28x+1)-\frac{1}{6}(x^{2}+3x+1)\ln x,\\
F_{\phi f\phi fbb}\left(x\right) & \equiv & 
\frac{1}{144x^{2}}\left(x-1\right)^{2}\left(3x^{3}+47x^{2}+11x-1\right)\nonumber \\
 &  & -\frac{1}{12}\left(x-1\right)\left(2x+3\right)\ln x.
\end{eqnarray}
In the above, besides the compensating factor from final state phase space
$c_{\rm PS_F}$, we have explicitly displayed the possible gauge degrees of 
freedom $c_{\rm G}$ and symmetry factor $c_{\rm S}$ from Wick's contraction 
when there are identical particles in the operator. 
The multiplicity factors for the operators \eqref{eq:before_EW_op}, 
\eqref{eq:after_EW_op1}--\eqref{eq:after_EW_op4} are listed 
in Table~\ref{tab:multiplicity_factors}.

From the operator of the type $\phi \phi f f f f$, the relevant 
reduced cross sections are
\begin{eqnarray}
\hat{\sigma}_{\phi\phi ffff}\left(x\right) & = &
c_{\rm G}\, c_{\rm S}\, c_{\rm PS_F} \frac{m_{\phi}^{8}}{\Lambda^{8}}
\beta\left(1,x^{-1},x^{-1}\right)\frac{1}{2^{11}\pi^{5}}\frac{x^{4}}{720}.\\
\hat{\sigma}_{\phi f\phi fff}\left(x\right) & = & 
c_{\rm G}\, c_{\rm S}\, c_{\rm PS_F} \frac{m_{\phi}^{8}}{\Lambda^{8}}
\beta\left(1,x^{-1},0\right)\frac{1}{2^{11}\pi^{5}}F_{\phi f\phi fff}\left(x\right),
\end{eqnarray}
where
\begin{eqnarray}
F_{\phi f\phi fff}\left(x\right) & = & 
\frac{\left(x-1\right)^{2}\left(4x^{4}+159x^{3}+239x^{2}+19x-1\right)}{1440x^{2}}\nonumber \\
 &  & -\frac{1}{24}\left(x-1\right)\left(x^{2}+4x+2\right)\ln x.
\end{eqnarray}

\begin{table}
\center
\begin{tabular}{|c||c|c|c|c|}
\hline 
 & $\boldsymbol{c_{{\rm G}}}$ & $\boldsymbol{c_{{\rm S}}}$ & $\boldsymbol{c_{{\rm PS_{F}}}}$ & $\boldsymbol{c_{{\rm PS_{I}}}}$\tabularnewline
\hline 
\hline 
$\boldsymbol{XX\bar{{\cal O}}^{\left(5\right)}}$ & $2\times2$ & $4\times4\times4$ & $\left(\frac{1}{2}\times\frac{1}{2},1\times\frac{1}{2}\right)$ & $\left(\frac{1}{2},1\right)$\tabularnewline
\hline 
$\boldsymbol{XX\bar{{\cal O}}^{\left(6\right){\rm I}}}$ & $3!\times2$ & $4$ & $\left(1,1\right)$ & $\left(\frac{1}{2},1\right)$\tabularnewline
\hline 
$\boldsymbol{XX\bar{{\cal O}}^{\left(6\right){\rm II}}}$ & $3!\times2$ & $4$ & $\left(1,1\right)$ & $\left(\frac{1}{2},1\right)$\tabularnewline
\hline 
$\boldsymbol{XX\bar{{\cal O}}^{\left(6\right){\rm III}}}$ & $3!\times2\times2$ & $4$ & $\left(1,1\right)$ & $\left(\frac{1}{2},1\right)$\tabularnewline
\hline 
$\boldsymbol{XX\bar{{\cal O}}^{\left(6\right){\rm IV}}}$ & $3!$ & $4$ & $\left(1,1\right)$ & $\left(\frac{1}{2},1\right)$\tabularnewline
\hline 
\end{tabular}
\caption{Multiplicity factors assuming that the operators only involve
the first family SM fermions (the family indices in the operators are dropped). 
For $c_{\rm PS_F}$ and $c_{\rm PS_I}$, 
the first entry is for scattering of the type $XX \to ijkl$ while 
the second entry is for $X i \to \bar X jkl$.
Here $X$ is either a complex scalar $\phi$ or a Dirac fermion $\psi$. 
\label{tab:multiplicity_factors}}
\end{table}

\subsection[Dark Matter as a Dirac Fermion $\psi$]{Dark Matter as a Dirac Fermion $\boldsymbol{\psi}$}

Here $\psi$ denotes a Dirac fermion with mass $m_\psi$ 
while $f$ and $b$ denote respectively a massless Weyl fermion and a complex scalar.  
From the operator of the type $\psi \psi f f b b$, 
the relevant reduced cross sections are
\begin{eqnarray}
\hat{\sigma}_{\psi\psi ffbb}\left(x\right) & = & 
c_{\rm G}\, c_{\rm S}\, c_{\rm PS_F} \frac{m_{\psi}^{8}}{\Lambda^{8}}
\beta\left(1,x^{-1},x^{-1}\right)\frac{1}{2^{11}\pi^{5}}\frac{(x-4)x^{3}}{36},\\
\hat{\sigma}_{\psi b\psi bff}\left(x\right) & = & 
c_{\rm G}\, c_{\rm S}\, c_{\rm PS_F}\frac{m_{\psi}^{8}}{\Lambda^{8}}
\beta\left(1,x^{-1},0\right)\frac{1}{2^{11}\pi^{5}}F_{\psi b\psi bff}\left(x\right),\\
\hat{\sigma}_{\psi f\psi fbb}\left(x\right) & = & 
c_{\rm G}\, c_{\rm S}\, c_{\rm PS_F}\frac{m_{\psi}^{8}}{\Lambda^{8}}
\beta\left(1,x^{-1},0\right)\frac{1}{2^{11}\pi^{5}}F_{\psi f\psi fbb}\left(x\right),
\end{eqnarray}
where $x \equiv s/m_\psi^2$ and
\begin{eqnarray}
F_{\psi b\psi bff}\left(x\right) & \equiv & 
\frac{1}{180 x^2} (x^2-1)\left(x^4 -24x^3 -374 x^2 -24x +1\right)  \nonumber \\
& &+ \frac{1}{3}(3x^2+8x+3)\ln x,\\ 
F_{\psi f\psi fbb}\left(x\right) & \equiv & 
\frac{1}{180 x^3} (x - 1)^2 (2x^5 -34x^4 -319x^3 -79x^2 +11x -1) \nonumber  \\  
& & + \frac{1}{3}(3x^2+x-4) \ln x .
\end{eqnarray}

From the operator of the type $\psi \psi f f f f$, 
the relevant reduced cross sections are
\begin{eqnarray}
\hat{\sigma}_{\psi\psi ffff}\left(x\right) & = & 
c_{\rm G}\, c_{\rm S}\, c_{\rm PS_F} \frac{m_{\psi}^{10}}{\Lambda^{10}}
\beta\left(1,x^{-1},x^{-1}\right)\frac{1}{2^{11}\pi^{5}}\frac{(x-4)x^{4}}{360},\\
\hat{\sigma}_{\psi f\psi fff}\left(x\right) & = & 
c_{\rm G}\, c_{\rm S}\, c_{\rm PS_F}\frac{m_{\psi}^{10}}{\Lambda^{10}}
\beta\left(1,x^{-1},0\right)\frac{1}{2^{11}\pi^{5}}F_{\psi f\psi fff}\left(x\right),
\end{eqnarray}
where
\begin{eqnarray}\nonumber
F_{\psi f\psi fff}\left(x\right) & = & 
\frac{1}{4320 x^3}(x-1)^2(5x^6-135x^5-3012x^4-4062x^3-387x^2+33x-2) \\
& & + \frac{7}{72}(x-1)(3x^2+10x+5)\ln x.
\end{eqnarray}

\bibliographystyle{JHEP}
\bibliography{biblio}

\end{document}